\RequirePackage{fix-cm}
\documentclass[twocolumn,epjc3]{svjour3}
\smartqed  
\RequirePackage{graphicx}
\RequirePackage{url}
\RequirePackage{lscape}
\RequirePackage{longtable}

\journalname{Eur. Phys. J. C}

\begin{document}

\title{Background assessment for the TREX Dark Matter experiment}


\author{
        J.~Castel\thanksref{addr1,addr2}
        \and S.~Cebri\'{a}n\thanksref{e1,addr1,addr2}
        \and I.~Coarasa\thanksref{addr1,addr2}
        \and T.~Dafni\thanksref{addr1,addr2}
        \and J.~Gal\'an\thanksref{addr1,addr2,addr3}
        \and F.J.~Iguaz\thanksref{addr1,addr2,addr4}
        \and I.G.~Irastorza\thanksref{addr1,addr2}
        \and G.~Luz\'o´n\thanksref{addr1,addr2}
        \and H.~Mirallas\thanksref{addr1,addr2}
        \and A.~Ortiz de Sol\'orzano\thanksref{addr1,addr2}
        \and E.~Ruiz-Ch\'oliz\thanksref{addr1,addr2}
        }

\thankstext{e1}{e-mail: scebrian@unizar.es}


\institute{Laboratorio de F\'{i}sica Nuclear y Astropart\'{i}culas,
Universidad de Zaragoza, Calle Pedro Cerbuna 12, 50009 Zaragoza,
Spain \label{addr1}
           \and
           Laboratorio Subterr\'{a}neo de Canfranc, Paseo de los Ayerbe s/n,
22880 Canfranc Estaci\'{o}n, Huesca, Spain\label{addr2}
           \and
           \emph{Present Address:} INPAC and Department of Physics and Astronomy, Shanghai Jiao Tong University, Shanghai Laboratory for Particle Physics and Cosmology, 200240 Shanghai, China\label{addr3}
           \and
           \emph{Present Address:} Synchrotron Soleil, BP 48, Saint-Aubin, 91192 Gif-sur-Yvette, France\label{addr4} }

\date{Received: date / Accepted: date}

\maketitle

\begin{abstract}
TREX-DM is conceived to look for low-mass Weakly
Interacting Massive Particles (WIMPs) using a gas Time
Projection Chamber equipped with Micromegas readout planes at
the Canfranc Underground Laboratory. The detector can hold in the active volume $\sim$20~l of
pressurized gas up to 10~bar, corresponding to 0.30~kg of Ar or 0.16~kg of Ne. The Micromegas are read with a
self-triggered acquisition, being thresholds below
0.4~keV (electron equivalent) at reach. A low background
level in the lowest energy region is another essential requirement. To assess the expected background, all the relevant sources have been considered, including the
measured fluxes of gamma radiation, muons and neutrons at the
Canfranc Laboratory, together with the activity of most
of the components used in the detector and ancillary systems,
obtained in a complete assay program. The background
contributions have been simulated by means of a dedicated
application based on Geant4 and a custom-made code for the detector response. The background model developed for the detector presently installed in Canfranc points to
levels from 1 to 10~counts keV$^{-1}$ kg$^{-1}$ d$^{-1}$
in the region of interest, making TREX-DM competitive in the search
for low-mass WIMPs. A roadmap to further decrease it down to 0.1~counts keV$^{-1}$ kg$^{-1}$ d$^{-1}$ is underway.
\end{abstract}

\section{Introduction}

Different detector technologies have been developed in the last
decades with the aim to directly detect dark matter particles which
could be pervading the galactic halo \cite{marrodan}. Looking
specifically for low mass Weakly Interacting Massive Particles
(WIMPs) requires the use of light elements as target, detectors with
very low energy threshold, well below 1~keV$_{ee}$\footnote{Electron
equivalent energy.}, and very low radioactive background. Results
from either new implementations of semiconductor detectors with
extremely low readout capacitance \cite{rescogent,resdamic,rescdex}
or re-oriented experiments focused on low threshold
\cite{rescdms,rescresst,resedelweiss,resxenon,darkside18,newsgresults} have already been presented.
Gas Time Projection Chambers (TPCs) equipped with Micromegas planes
have excellent features to fulfill these requirements. TREX-DM (TPC
for Rare Event eXperiments-Dark Matter) \cite{iguaz16,irastorza16}
is a Micromegas-read High Pressure TPC for low mass WIMP searches
using Ar or Ne mixtures, not focused on directionality. The detector
was built and operated at surface in the University of Zaragoza as
proof of concept. The experiment was approved by the Canfranc
Underground Laboratory (LSC) in Spain and the whole detector was moved underground in 2018. The data taking is expected to
start using Ne once the commissioning underground along 2019 is completely finished.

The Micromegas are consolidated readout structures; a micro-mesh is
suspended over a pixelated anode plane, forming a thin gap where
charge amplification takes place. Detectable signals in the anode
and the mesh are generated. Different technologies have been built:
{\it bulk} Micromegas have the readout plane and the mesh all in one and
{\it microbulk} Micromegas are in addition more homogeneous and radiopure
\cite{microbulk}. They offer important advantages for rare event
detection \cite{irastorza16,irastorza16dbd}: the possibility of
scaling-up, topological information to discriminate backgrounds from
the expected signal (just a few microns track for dark matter
particles, giving a point-like event) and low intrinsic
radioactivity as they are made out of kapton and copper, potentially
very clean.

Operating deep underground at ultra-low background conditions is a must in experiments looking for rare events like the direct
detection of WIMPs. In this kind of experiments, the construction of
reliable background models, based on an accurate assay of background
sources and on a careful computation of their contribution to the
experiment, is essential; they provide guidance and constraints for
design and allow robust estimations of the experiment sensitivity
(some recent examples can be found at \cite{xenon100,zeplin,edelweiss,gerda,lux,exobkg,anais,cuorebkg,cosine}). The
preliminary background model of TREX-DM for operation at LSC
presented in \cite{iguaz16,Aznartaup:2017} has now been completed
and updated, including as inputs the activities from a dedicated
material screening program together with the measured fluxes of
different backgrounds at LSC (gamma-rays, neutrons and muons).

The structure of the article is the following. The detector set-up
and its performance are presented in section~\ref{exp}.
Section~\ref{sim} describes the simulations carried out. Then, the
results obtained in the material radioassay campaign are detailed in
section~\ref{rad}. The estimates of the contribution of each one of
the background components considered are shown in section~\ref{bac}.
Finally, the corresponding sensitivity for WIMP direct detection and
conclusions are discussed in sections~\ref{sen} and \ref{con}.

\section{Experimental set-up}
\label{exp}

The TREX-DM detector, as built and operated at the University of
Zaragoza, was described in detail in \cite{iguaz16}; in the set-up at LSC some modifications have been implemented: operation is made with non-flammable gas mixtures, microbulk (instead of bulk) Micromegas are being used read by a new AGET-based DAQ system and a full detector shielding is in place.

\begin{figure}
\centering
\includegraphics[width=0.25\textwidth]{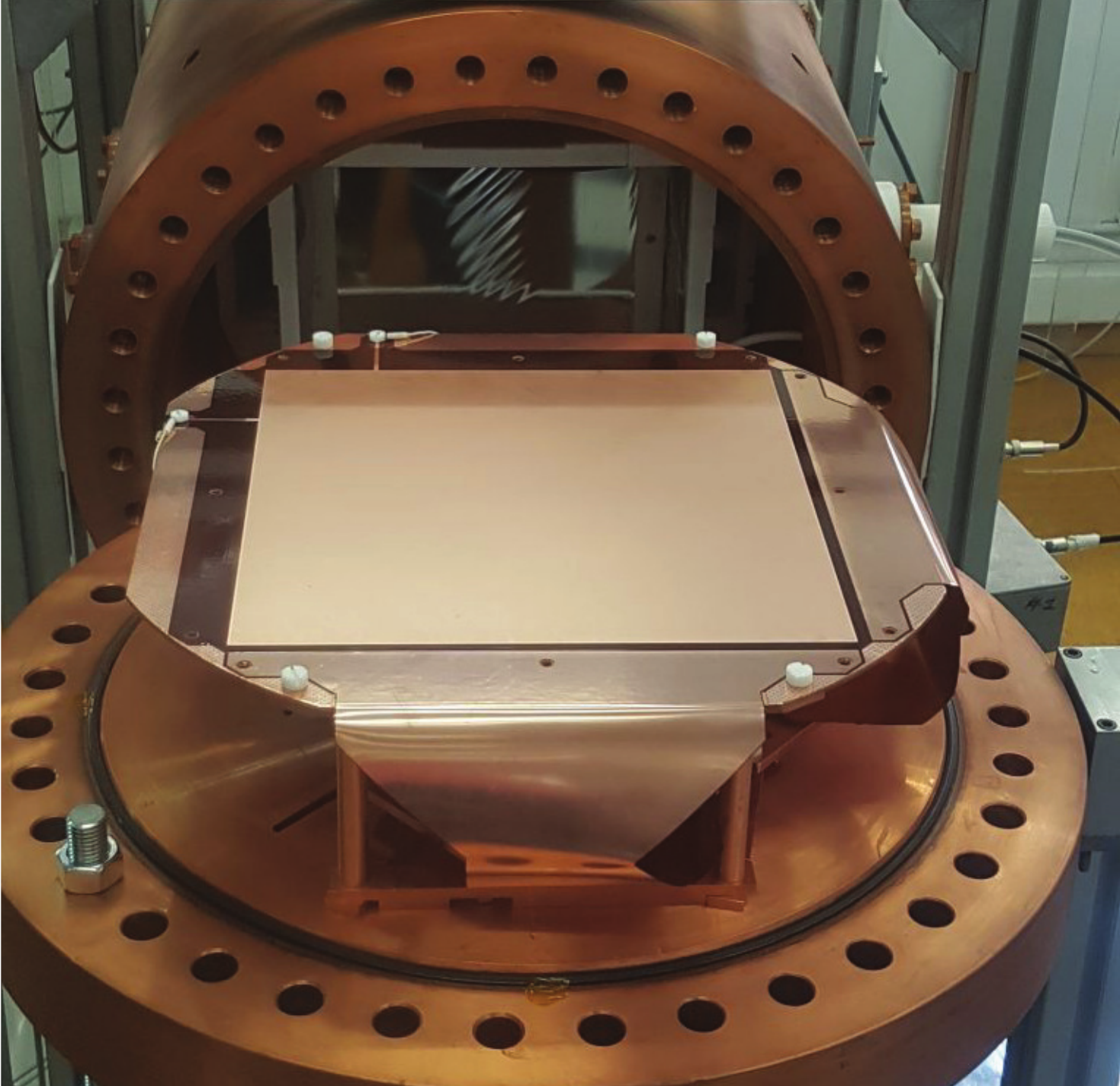}\\ 
\includegraphics[width=0.25\textwidth]{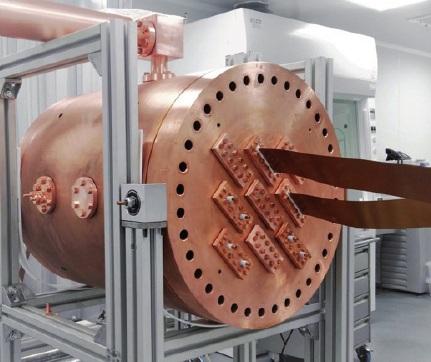}\\
\includegraphics[width=0.25\textwidth]{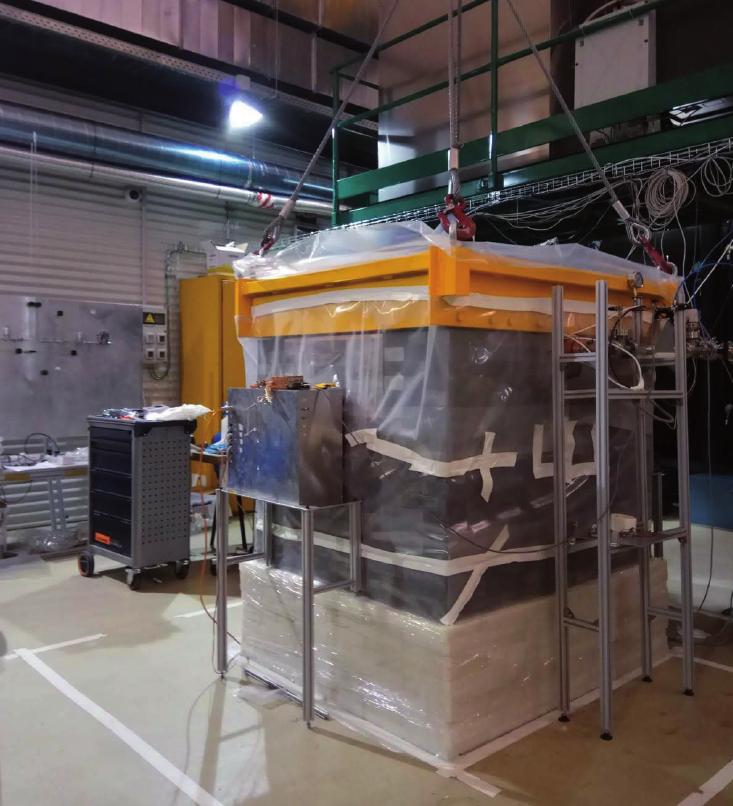}
\caption{Pictures of the Microbulk Micromegas of TREX-DM, the largest ever
fabricated, being tested (top), the copper vessel at the clean room of LSC during mounting (middle) and the set-up with the lead shielding at hall A of LSC (bottom).} \label{picture}
\end{figure}

Two active
volumes (19$\times$25$\times$25 cm$^{3}$ each) are separated by a
central cathode made of mylar inside a 6-cm-thick copper vessel,
designed and certified as a pressure equipment to hold up to 11~bar(a) or 10~barg (see figure~\ref{picture}, middle). The field cage, made of kapton and
copper using 42 Finechem resistors, is covered by teflon.
Two microbulk Micromegas fabricated at CERN are used; they are the largest area single microbulk readout produced so far, with an active area of 25$\times$25~cm$^{2}$
(see figure~\ref{picture}, top). Flat cables take out signals
from strips and connect to the interface cards out of the vessel. The connections at both sides of the flat cables are now made through special silicone-based connectors (Zebra Gold 8000C from Fujipoly),
checked to be more radiopure than the connectors firstly used (see section~\ref{rad}). Signals from
2$\times$256 strips (with $\sim$1~mm pitch) at each side are
digitized for tracking in a 10.2~$\mu$s window (512~samples, 50~MHz
sampling rate). 3D track reconstruction is possible, using the position of hit
strips to determine X and Y coordinates, and electron drift time
information to get the Z coordinate. The AGET-based DAQ system
consists of four front-end (FEC-AGET)
cards, containing 4 AGET chips each for sampling of the pulses, four
mezzanine (FEMINOS) cards, subtracting pedestals, and a trigger
module (TCM), connected by Ethernet cables. It provides
self-triggered data acquisition allowing to have a trigger generated individually by each single strip signal; as capacitance of strips is about a factor 30 lower than that of the mesh, the use of AGET electronics was considered in order to
improve the energy threshold in comparison to the first operation using AFTER electronics with trigger taken from the mesh signal \cite{iguaz16}. Two Faraday cages, one at each end,
have been installed housing the interface cards, FEC-AGET and FEMINOS
cards (see figure~\ref{picture}, bottom).

\begin{figure}
\centering
\includegraphics[width=0.5\textwidth]{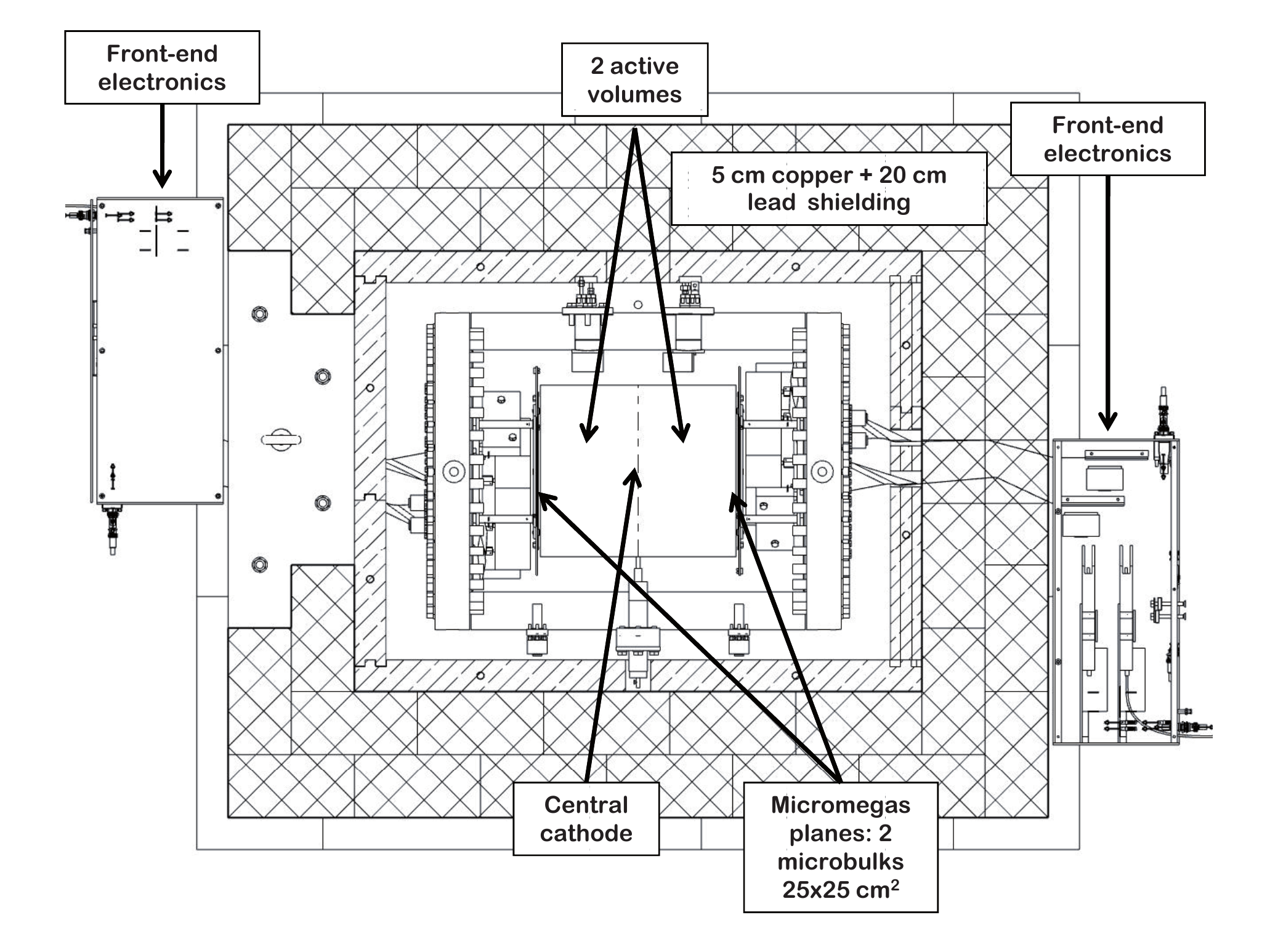}
\includegraphics[width=0.5\textwidth]{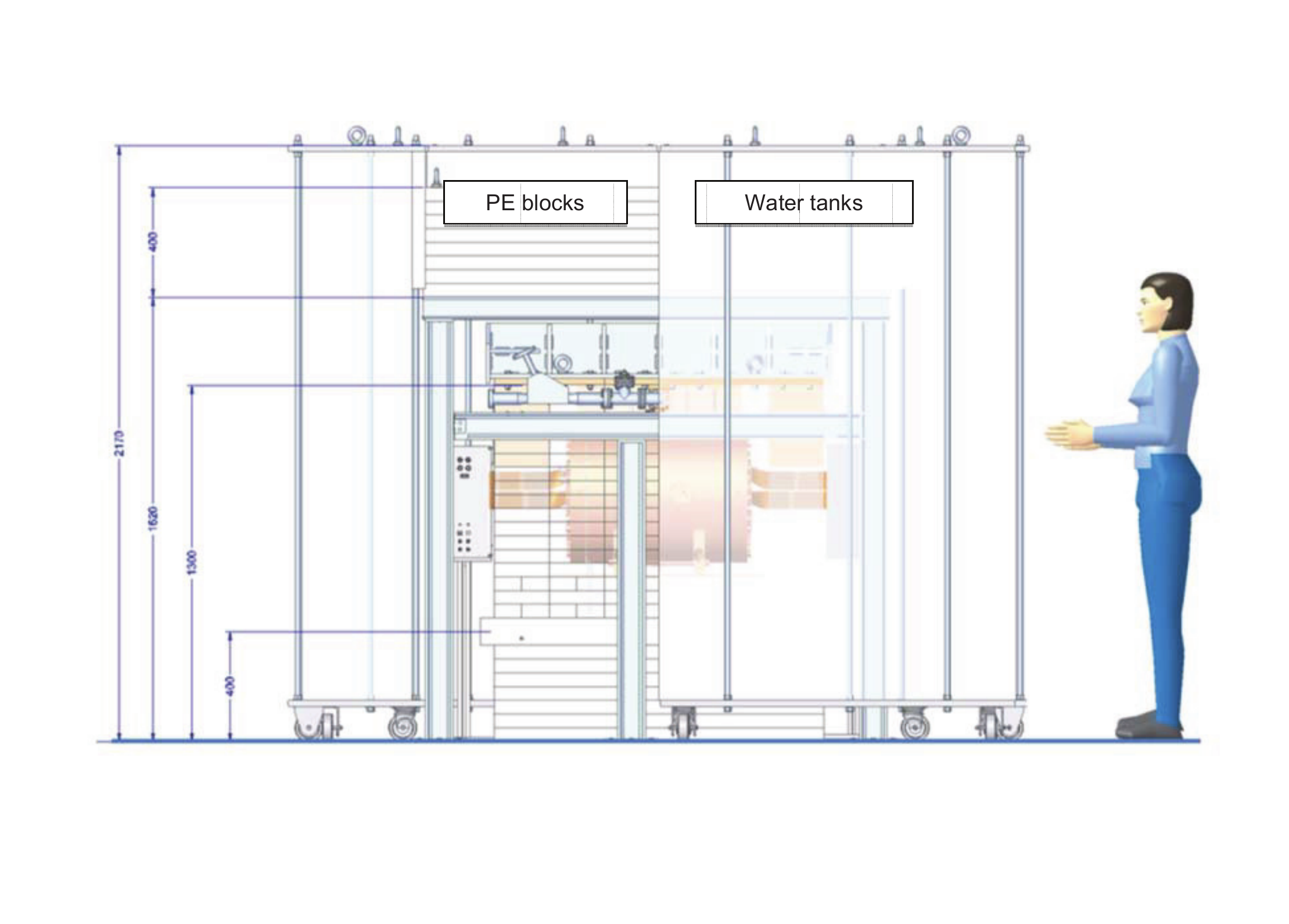}
\caption{Top view of the design of the copper and lead shielding (top) and side view of the moderator
shielding (bottom) for TREX-DM at LSC.} \label{shielding}
\end{figure}

As shown in figure~\ref{shielding}, a complete shielding consisting
of 5~cm of copper, 20~cm of low activity lead and 40~cm of neutron
moderator (water tanks and polyethylene blocks) covers the
detector at LSC, having in addition a Rn-free atmosphere
inside shielding. The present TREX-DM set-up installed at the hall A of the Canfranc laboratory, placed at a depth of
2450 m.w.e., is shown in figure~\ref{picture}, bottom, including the lead shielding.

The gas system of TREX-DM has been designed for non-flammable gases,
which is accomplished adding 1\% of isobutane to argon and 2\% in
the case of neon. This simplified installation underground. It
consists of a recirculation part and a purification branch and it can run in two operation modes: in
open loop (for commissioning tasks) or in recirculation through a
purifier branch (nominal working condition). The gas system installation has been completed
and certified by an authorized body to operate at high pressure (up to 11~bar). The slow control
system, based on a net of Raspberry PI cards, each of them
monitoring or controlling one experimental component (like gas
pressure and temperature, voltages or gas flow), is
ready and in operation.

\begin{figure}
\begin{center}
\includegraphics[width=0.5\textwidth]{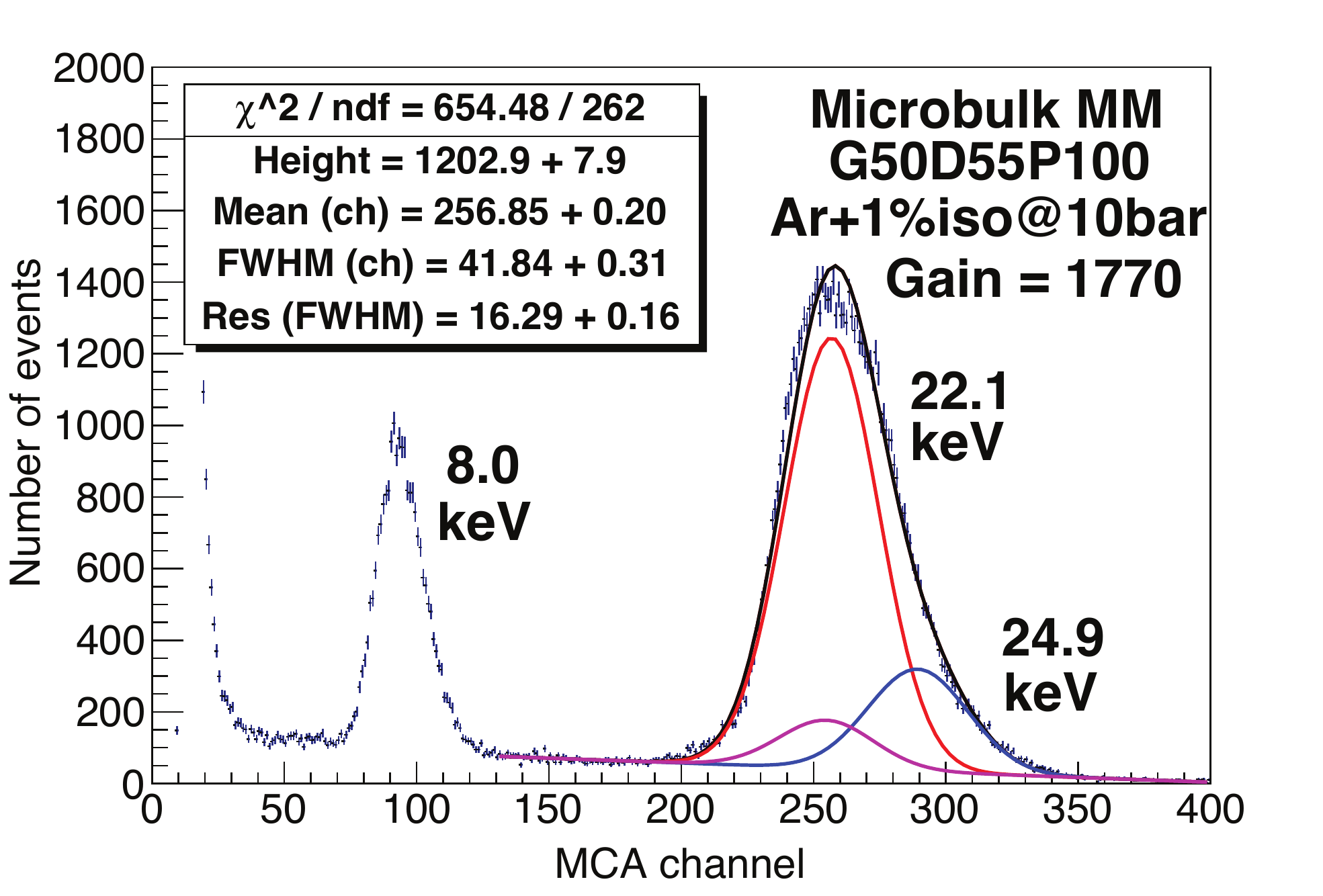}
\includegraphics[width=0.5\textwidth]{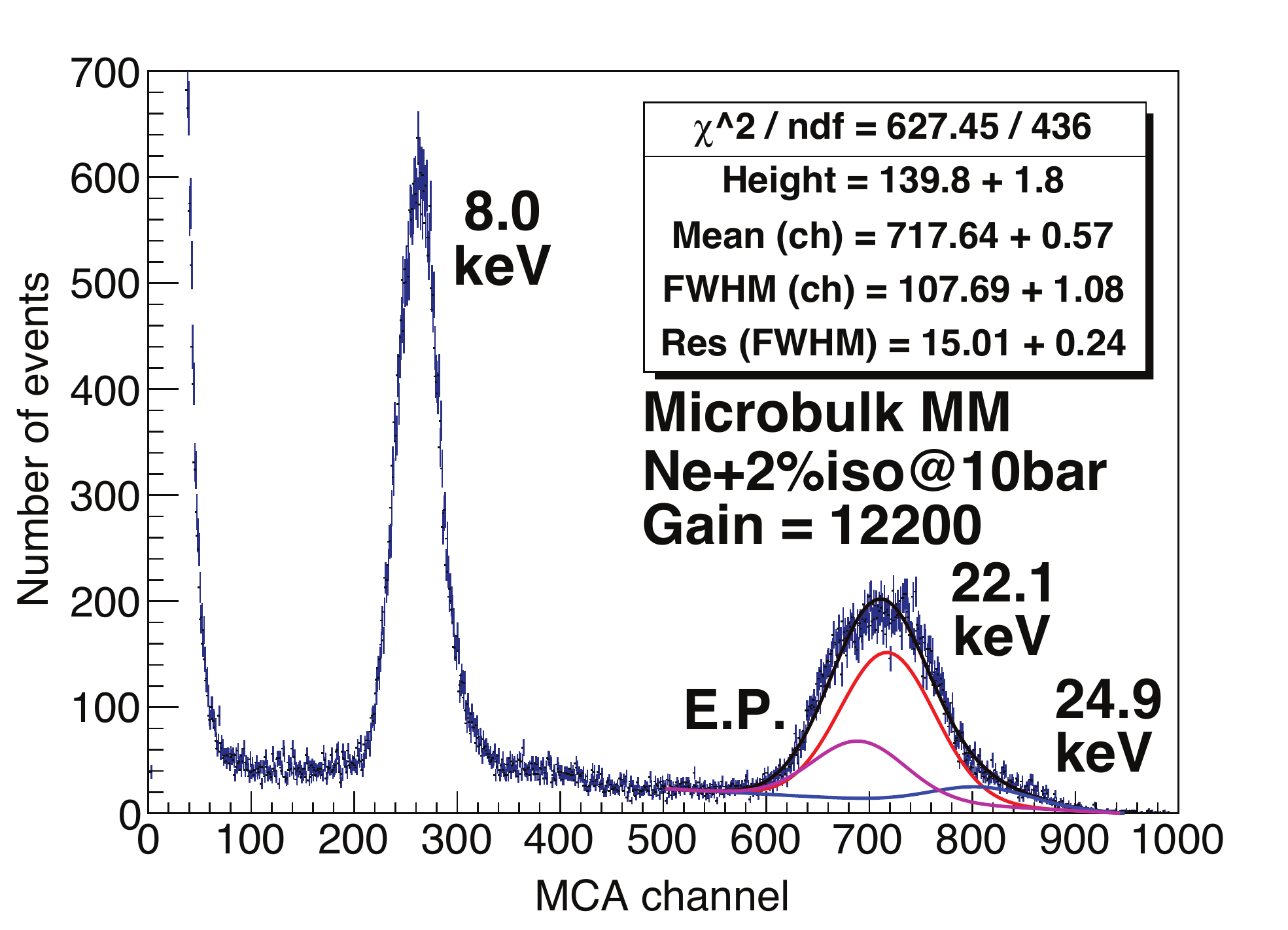}
\caption{Energy spectrum obtained in the characterization of microbulk Micromegas with Ar (top) and Ne (bottom) mixtures using a
$^{109}$Cd source at 10~bar. The results of a multi-Gaussian fit for
estimating the FWHM at 22.1~keV are shown. The fitted peak is generated by the
K$\alpha$ (22.1~keV) and K$\beta$ (24.9~keV) X-ray lines produced by the source, as well as their corresponding escape peaks. The copper K-fluorescence induced by the source at the copper components close to the gas volume is clearly identified at 8~keV.} \label{spcs}
\end{center}
\end{figure}

\begin{figure}
\begin{center}
\includegraphics[width=0.5\textwidth]{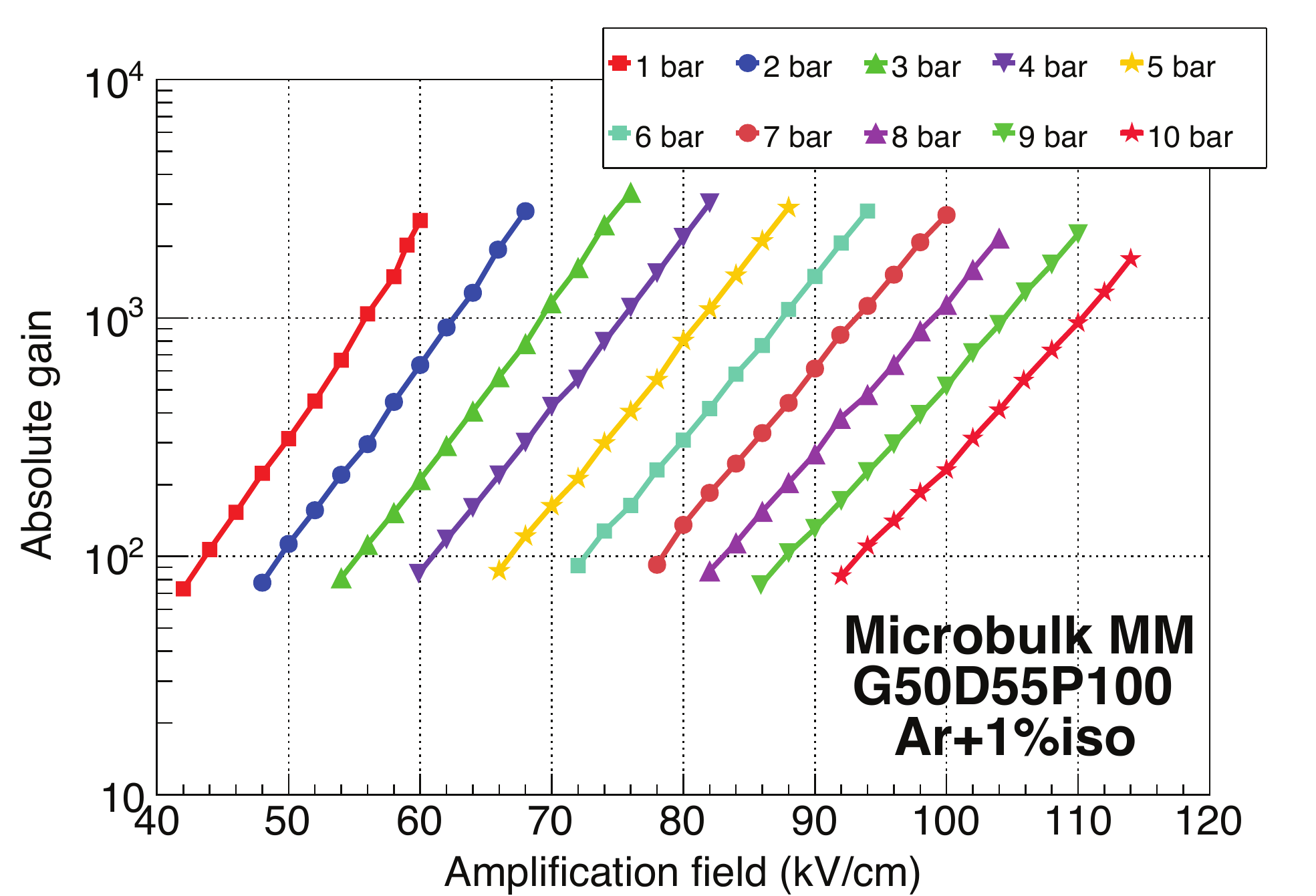}
\includegraphics[width=0.5\textwidth]{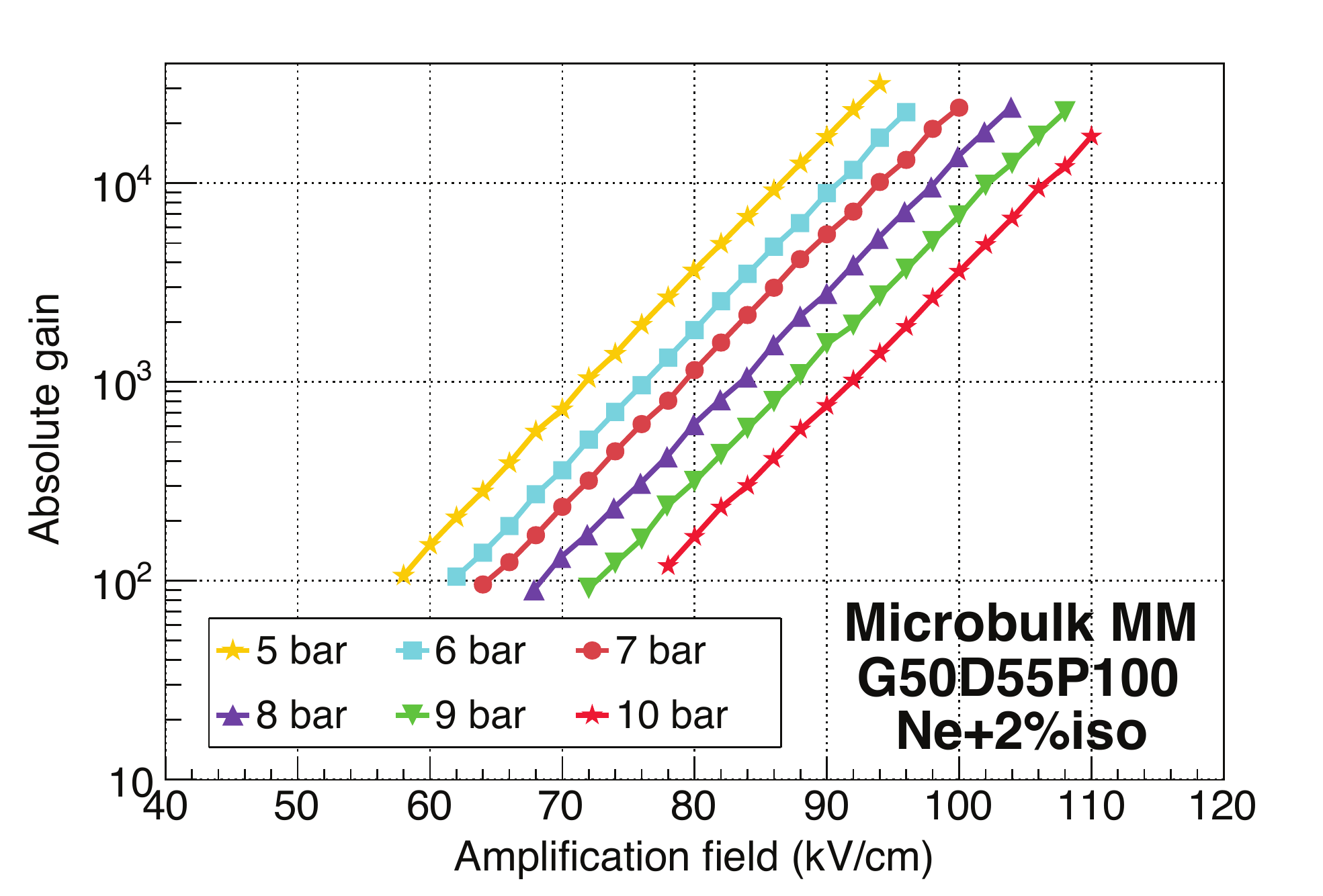}
\caption{Gain curves obtained in the characterization of microbulk Micromegas with Ar (top) and Ne (bottom) mixtures using a
$^{109}$Cd source. Curves for different pressures up to 10~bar are shown.} \label{gains}
\end{center}
\end{figure}

First results from the commissioning phase of TREX-DM on surface using bulk Micromegas
were shown in \cite{iguaz16} for different gas mixtures and pressures. Microbulk Micromegas
were characterized afterwards using the mixtures finally considered for the underground operation, Ar+1\%iC$_{4}$H$_{10}$ and
Ne+2\%iC$_{4}$H$_{10}$, in order to check that the required performance was achieved also in these conditions \cite{iguaz18}. These tests were made at 1-10~bar using a $^{109}$Cd
source in a smaller set-up at the University of Zaragoza (also used in \cite{mmxe}) without shielding. Figure~\ref{spcs} presents the energy spectra registered at 10 bar; energy resolution has shown some degradation with pressure,
being the FWHM at 10~bar 16(15)\% for Ar(Ne) at 22.1~keV. It must be noted that neither the background level nor the energy threshold in these measurements were representative of those achievable at TREX-DM in LSC. As shown in figure~\ref{gains}, an excellent behavior has been registered for gain, with maximum values above 10$^{3}$(10$^{4}$) in Ar(Ne) for all
pressures, which is very important for achieving low energy
thresholds. In principle, a very low threshold is possible thanks to
the intrinsic amplification in gas but the readout area,
the sensor capacitance and the electronic noise set in practice the effective threshold in this type of detectors. For TREX-DM, the strip capacitance, including the contributions
of the flat cables and the interface cards, is $\sim$0.2~nF. The TREX-DM nominal (conservative) aim for effective threshold is 100~eV$_{ee}$
(400~eV$_{ee}$). The very first results from the commissioning data taken in Canfranc also using microbulk Micromegas with atmospheric Ar and with Ne confirm that this expectation is at reach.

At the end of 2018, the detector was already fully equipped (as described in this section) and installed inside its copper and lead castle at LSC. For the first commissioning data, the chamber was filled with atmospheric Ar+1\%iC$_{4}$H$_{10}$ and operated at 1-1.5 bar. Calibration measurements made with a $^{109}$Cd source indicated that the detector operates as expected in terms of gain and spectral and spatial distribution of events. Few-days long background runs with the expected performance were carried out, being the measured spectra dominated by $^{39}$Ar. In the beginning of 2019, the
gas changed from the test option to the baseline mixture Ne+2\%iC$_{4}$H$_{10}$ in recirculation and operation at higher pressures is underway.
Data analysis is ongoing to produce quantitative statements on e.g. threshold level and background level.  A detailed description of the performance of the detector will be presented in a dedicated technical publication. 


\section{Simulation}
\label{sim}

\begin{figure}
\centering
 \includegraphics[width=0.5\textwidth]{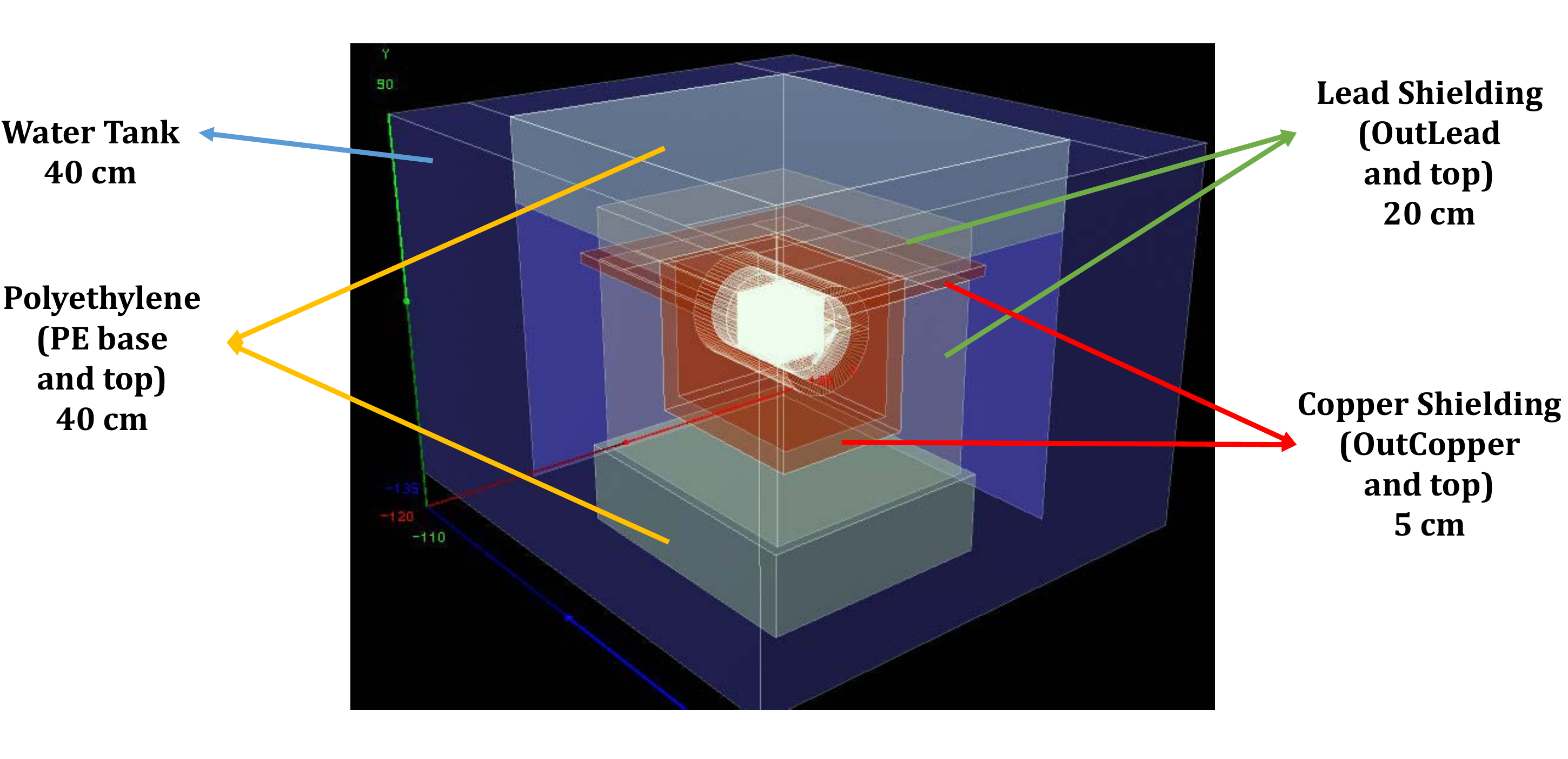}
 \includegraphics[width=0.5\textwidth]{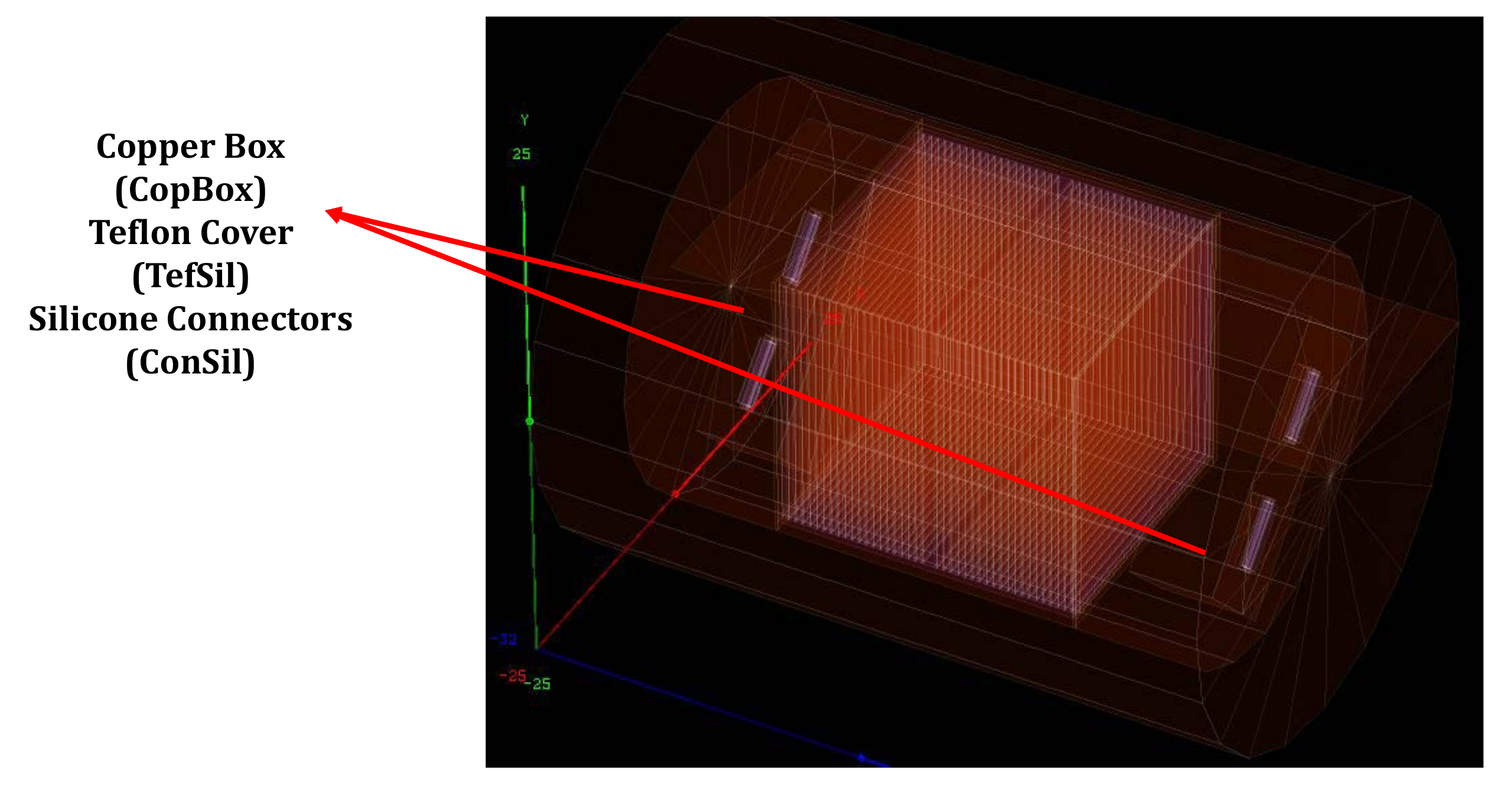}
\caption{Views of the geometry of TREX-DM implemented in the GEANT4 simulation code, showing the whole shielding (top) and the copper vessel (bottom).}
 \label{geometry}
\end{figure}

The complete simulation of the detector response to build the TREX-DM
background model is based on RestG4, a package integrating Geant4 \cite{geant4} and a
custom-made code called REST~\cite{restsoft}, also used in \cite{panda}. Geant4 version 4.10 has been considered. The Geant4 Radioactive Decay Module is used for the event generation. The low energy models based on Livermore data libraries have been implemented for interactions of alpha, beta and gamma particles. These models are accurate for energies between 250~eV and 100~GeV and can be applied down to 100~eV with a reduced accuracy \cite{geant42}. The NeutronHP model has been used for neutrons.
A detailed geometry of the set-up has been implemented based on the Geometry Description Markup Language (GDML) format (see figure~\ref{geometry}), including the whole shielding (with copper, lead and neutron moderator) and details of the copper vessel, field cage, cathode, cables and connectors and Micromegas readouts. The ROOT-based code simulates also the electron generation in gas, diffusion effects, charge amplification at Micromegas and signal generation. The resulting data from the simulation chain have the same REST format as the DAQ data in order to share analysis tools, including those for discriminating point-like events from complex topologies.

Simulations to build the TREX-DM background model have been run for the two gas mixtures (Ar+1\%iC$_{4}$H$_{10}$ and Ne+2\%iC$_{4}$H$_{10}$) at 10~bar, corresponding to a total active mass of 0.30~kg for Ar and 0.16~kg for Ne. A successful validation of the simulation environment against experimental data from calibration measurements has been made \cite{iguaz16}; in particular, the shape of the energy spectra and the distribution of the analysis observables measured were checked to be properly reproduced by simulation when irradiating the active volume with a $^{109}$Cd source. The event-by-event signal identification and background rejection applied is based on the analysis of the event topology. Firstly, just one-cluster events are selected. This selection criterion hardly reduces the rate a 0.1\% since in the region of interest almost all the events are single-site. The second condition is to be farther than 2~cm from the lateral faces of the prism defined by the active volume; i.e., a planar fiducial cut. Simulations show an accumulation of events near the border of the readout plane due to photons produced in interactions in surrounding materials (field cage, cathode frame, \dots), which disappears after imposing this 2-cm-veto-area. Reduction factors in this case depend very much on the background sources, but have mean values of $\sim$40\%. As our detector lies on one of its sides, this veto also eliminates muons traversing directly the detector. The active mass of the detector to obtain the counting rates per mass unit is that of the fiducial volume. So far, the analysis has been quite simple and more complex discrimination algorithms, based on the analysis of current calibration runs, could help to further decrease background levels.

\section{Measurements of material radiopurity}
\label{rad}

An exhaustive material radioassay campaign for TREX-DM was
undertaken a few years ago
\cite{iguaz16,irastorza16dbd,jinstpaquito}, as made in other
experiments in the context of rare event searches (see for instance
recent results in
\cite{xenon,exo,majorana,next1,next2,next3,coach}). It has allowed,
on one side, to design and construct the detector and shielding
according to the radiopurity specifications and, on the other, to
provide inputs to build the experiment background model. The
material screening program is mainly based on germanium gamma-ray
spectrometry carried out deep underground in Canfranc, but
complemented by other techniques like GDMS and ICPMS and by
measurements using the BiPo-3 detector at LSC. In this section, the
techniques employed for the radiopurity measurements are summarized
and the relevant results shown and discussed.

\subsection{Techniques}

Most of the germanium measurements have been made using a $\sim$1~kg
ultra-low background detector of the University of Zaragoza (named
Paquito) operated at the hall E of the LSC. This detector has been
used for radiopurity measurements at Canfranc for many years
(details on the features and performance of the detector can be
found in \cite{jinstpaquito,mmradiopurity}).
Some of the $\sim$2~kg close-end coaxial HPGe detectors
of the Radiopurity Service of LSC \cite{bandac17} have been used for
some measurements too. Activities of different sub-series in the
natural chains of $^{238}$U, $^{232}$Th and $^{235}$U as well as of
common primordial, cosmogenic or anthropogenic radionuclides like
$^{40}$K, $^{60}$Co and $^{137}$Cs are typically evaluated. The
detection efficiency is determined by Monte Carlo simulations based
on Geant4 for each sample, validated with a $^{152}$Eu reference
source \cite{jinstpaquito}; a conservative overall uncertainty on
the deduced efficiency is properly propagated to the final results.

Glow Discharge Mass Spectrometry (GDMS) has been performed by Evans
Analytical Group in France, providing concentrations of U, Th and K.
In addition, thanks to the collaboration of LSC, Inductively Coupled
Plasma Mass Spectrometry (ICPMS) analysis carried out at the
Laboratori Nazionali del Gran Sasso (LNGS) \cite{icpmsgs} has been possible for
some TREX-DM samples quantifying the U and Th concentrations. It
must be noted that when having no information on daughter
radionuclides in the chains, a possible disequilibrium cannot be
detected.

Taking advantage of the ``foil'' geometry of some samples, very
sensitive measurements have been made in the BiPo-3 detector
\cite{bipo3}, in operation at LSC. This detector has been developed
by the SuperNEMO collaboration and is able to measure extremely low
levels, down to a few $\mu$Bq/kg, of $^{208}$Tl and $^{214}$Bi
radioactivity in very thin samples (below 200~$\mu$m thick) by
registering the delayed coincidence between electrons and alpha
particles occurring in the BiPo events. These measurements can be
translated into contamination of natural U and Th chains if secular
equilibrium is assumed.

\subsection{Results}

A large amount of materials and components related to Micromegas
readout planes and the whole set-up (the gas vessel, the field cage,
the radiation shielding or the electronic acquisition system) has
been taken into consideration in the screening program. Massive
elements and those in contact with the sensitive volume of the
detector are in principle the most relevant. Table~\ref{res}
reproduces, for the sake of completeness, the previous activity
results obtained for the components finally used in the TREX-DM
set-up and presents all the new ones; reported errors include both
statistical and efficiency uncertainties.

\subsubsection{Shielding and vessel}
\label{radves}
Many samples from different suppliers of lead, used for shielding,
and copper, used for mechanical and electric components (like
Micromegas plates, cathodes, HV feedthroughs or field cage rings),
have been analyzed \cite{iguaz16}. Results are available for the lead bricks
used in the shielding, provided by Mifer (\#1 of
table~\ref{res}) and by LSC (from lead coming from the OPERA
experiment \cite{lrt2015}). Very stringent upper limits were also
obtained by GDMS for Oxygen Free Electronic (OFE, C10100) copper
from the Luvata company and the used Electrolytic Tough Pitch (ETP,
C11000) copper from Sanmetal was measured too (\#2-3 of
table~\ref{res}).

A sample of the copper provided by Sanmetal and used at the
vessel of TREX-DM, having the same origin and history of exposure to
cosmic rays on surface as the vessel itself, has been additionally
screened using one of the germanium detectors of the LSC Radiopurity
Service with the aim to evaluate the cosmogenic activation induced
in the vessel and then to assess its suitability for a low
background measurement at LSC. The measured $^{60}$Co activity (\#4
of table \ref{res}) is in very good agreement with expectations for
production rates of $\sim$50~kg$^{-1}$d$^{-1}$ $=$0.58~mBq/kg
(corresponding to the order of different direct measurements and
calculations from the literature \cite{ijmpacosmogenia}) and an
exposure time of $\tau /2=$3.80~y. The activity of other isotopes
with half-lives shorter than $^{60}$Co has also been evaluated:
(0.35$\pm$0.07)~mBq/kg of $^{58}$Co, $<$0.81~mBq/kg of $^{57}$Co and
$<$0.29~mBq/kg of $^{54}$Mn. Thanks to the low energy threshold of the
GeOroel detector, it was possible in this measurement to observe the
46.5~keV gamma line from $^{210}$Pb and to derive an upper limit on
its activity (considering the efficiency for surface emissions) as
$<$0.32~mBq/cm$^{2}$.

Two samples of tube intended to be used at the calibration system of
TREX-DM, traversing shielding and vessel have been screened (\#5-6 of
table~\ref{res}). One was made of PFA (PerFluoroAlkoxy polymer)
produced by Emtecnik and the other was made of PTFE supplied by
Tecnyfluor. Upper limits have been set for all the common
radioisotopes for both tubes and therefore any of them can be used.

\subsubsection{Field cage}
Material and components to be used inside the gas vessel, mainly
related to the field cage, were screened to select proper teflon,
resistors or adhesives.

The monolayer Printed Circuit Board (PCB) made of kapton and copper
for the field cage, supplied by LabCircuits, was found to have good
radiopurity \cite{iguaz16}, but since the upper limits set on the
activity were too high for the required sensitivity, a new sample
with larger surface was measured afterwards (\#7 of
table~\ref{res}), providing a reduction of at least a factor 10 in
the relevant isotopes.

For the epoxy resin Hysol RE2039 from Henkel
no contaminant could be quantified \cite{iguaz16,jinstpaquito} (\#8
of table~\ref{res}) and it is being used for gluing.

Surface Mount Device (SMD) resistors supplied by Finechem showed
lower activity than other equivalent units
\cite{iguaz16,jinstpaquito} (\#9 of table~\ref{res}) and therefore
were chosen for the field cage.

A sample of the 3.5-$\mu$m-thick mylar sheet from Goodfellow used at
the cathode of TREX-DM has been screened (\#10 of table~\ref{res}),
setting upper limits for all the common radiosotopes.

The possibility of using 3D printing to produce easily some of the
mechanical components of the field cage of TREX-DM, instead of
fabricating them using commercial teflon, was considered and the
possible impact on background evaluated. A sample made of
PA2200\footnote{PolyAmide white or polyamide 12 or nylon 12.}
produced by a 3D printer at the Centro Nacional de Microelectronica
(CNM), Barcelona, was first screened using a germanium detector
(\#11 of table~\ref{res}) setting upper limits for all the usual
radiosotopes. In an attempt to reduce the available limits on
activity for several plastic samples, which would give a
non-negligible contribution to the TREX-DM background model, an
ICPMS analysis was carried out at LNGS for three samples. For all of
them the mineralization was performed using the dry ashing technique
and the uncertainty of the measurement can be estimated as 30\% of
the given values. For teflon from Sanmetal, the
ICPMS limits set for the activity of $^{232}$Th and $^{238}$U chains
(\#13 of table~\ref{res}) improve by about two orders of magnitude
those obtained using the Paquito detector at LSC
(\cite{iguaz16,jinstpaquito}). Another type of teflon, extruded PTFE
(Gore GR gasketing), was also analyzed, pointing to a similar
radiopurity (\#14 of table~\ref{res}). For a small piece of the
nylon PA2200 from 3D printing, previously screened at a germanium
detector setting upper limits for all common radioisotopes, both U
and Th concentrations could be quantified. As shown in \#12 of
table~\ref{res}, the corresponding $^{232}$Th activity is compatible
with the upper limits set using germanium detectors, but being of
the same order. For $^{238}$U, the activity value from ICPMS is
compatible with that for the upper part of the chain from germanium
(the actual comparable result between both techniques) but higher
that than for the lower part. Therefore, following these radiopurity
results and given the important contribution expected from a nylon
PA2200 field cage in the background model, its use has been
disregarded and teflon has been employed instead.

\subsubsection{Electronics}
Electronic connectors made of Liquid Crystal Polymer (LCP) have
shown unacceptable activities of at least several mBq/pc for
isotopes in $^{232}$Th and the lower part of $^{238}$U chains and
for $^{40}$K \cite{iguaz16,jinstpaquito,next1,next2}. Three
different types of silicone-based connectors supplied by Fujipoly
have been screened: Gold 8000 connectors type C, units from Zebra
Series 5002 SZ100 made of silver, and connectors from Zebra Series
2004 CZ418 made of carbon (\#15-17 of table~\ref{res}); although
activities of $^{40}$K, $^{232}$Th and the lower part of the
$^{238}$U chain have been quantified in all cases, much lower values
than in LCP connectors have been obtained. For silver connectors,
activity from $^{208m}$Ag (T$_{1/2}$=428~y, (0.38$\pm$0.05)~mBq/pc)
and $^{210m}$Ag (T$_{1/2}$=249.78~d, (1.88$\pm$0.12)~mBq/pc) has
been assessed too. Due to these results, Gold Zebra 8000C connectors
were selected for TREX-DM; a new screening of the units to be
actually used has been made (\#18 of table~\ref{res}) confirming
that in comparison with the Samtec LCP connectors firstly used in
the set-up at the University of Zaragoza \cite{jinstpaquito}, the
silicone ones of the present set-up have about a factor 32 less
activity of $^{232}$Th and reduction is about 3.3 for $^{226}$Ra and
2.3 for $^{40}$K. Recently, very promising results have been found in the screening of new connectors produced by Samtec (Ref. ZA8H-24-0.33-Z-07) (\#19 of table~\ref{res}), as only upper limits have been set for all the common radioisotopes, being more than one order of magnitude lower than the quantified activities in the Fujipoly ones. Comparing to the firstly used Samtec connectors, the reduction is around two orders of magnitude. This has been possible due to the very reduced mass of each unit (0.087 g/unit) and the use of only kapton and copper as base materials. This is an important result to take into consideration in future upgrades of the detector.

Very radiopure, flexible, flat cables made of kapton and copper have
been developed in collaboration with Somacis, performing a
careful selection of the materials included and avoiding glass
fiber-reinforced materials at base plates. After the screening of
several cable designs \cite{iguaz16}, the good results obtained for
the final one (\#20 of table~\ref{res}) allow to envisage the use of
these materials also at Micromegas production. Several kinds of high
voltage or signal cables have been analyzed too \cite{iguaz16}.
Screened cables from Druflon Electronics (to connect the field cage
last ring to HV feedthrough) and coaxial low noise cable from Axon
Cable S.A.S. (to extract the mesh signal from the vessel) are being
used. Only $^{40}$K activity was quantified for the two cables
(\#21-22 of table \ref{res}), made basically of copper and teflon,
showing much better radiopurity than typical RG58 coaxial cables.

Different materials can be taken into consideration for PCBs and
samples of FR4, ceramic-filled PTFE composite and cuflon were
screened \cite{iguaz16}. The first ones presented very high
activities for the natural chains and $^{40}$K, precluding its use.
Good radiopurity was found for cuflon from Crane Polyflon; however,
its application for Micromegas has been disregarded due to the
difficulty to fix the mesh and also because bonding films to prepare
multilayer PCBs have been shown to have unacceptable activity
\cite{next2}. Following the measurements \#7 and \#20 of
table~\ref{res}, kapton-copper boards seem to be the best option.

One of the electronic boards used at TREX-DM (with approximate
surface 14$\times$25~cm$^{2}$) has been directly screened (\#23 of
table~\ref{res}). Values of specific activity obtained for
$^{232}$Th and $^{238}$U chains are a factor 2-3 larger than those
measured in a raw PCB made mainly of FR4 from Somacis company
\cite{iguaz16}, which seems to point out to an additional relevant
source of activity in the electronic components of the board. A
sample of non-functional AGET chips (2.74~g/pc) provided by CEA
Saclay has been screened too. Following the results in \#24 of
table~\ref{res}, each chip has a few tenths of mBq of $^{40}$K and
of the isotopes of the $^{232}$Th and $^{238}$U chains. These
non-negligible quantified activities in the electronic components
should not pose a problem since they are located outside the copper
and lead shielding of the DAQ system. As part of the common effort to
develop radiopure electronic components in collaboration with CEA
Saclay, a sample of four units of AGET chips with a different
ceramic cover provided by CEA was analyzed too (\#25 of
table~\ref{res}). Specific activities of the order of tens of Bq/kg
have been measured for the isotopes of the $^{232}$Th and $^{238}$U
chains; the corresponding activities per unit are from two to three
orders of magnitude higher in comparison to those previously
obtained for AGET chips with plastic cover. Therefore, the use of
this type of ceramic packaging must be avoided. A large sample of
naked chips has been screened too at the Modane Laboratory
confirming a very good radiopurity.

\subsubsection{Micromegas}
\label{secmicro}
The radiopurity of Micromegas readout planes was first analyzed in
depth in \cite{mmradiopurity}. On the one hand, two samples (\#26-27
of table~\ref{res}) were part of fully functional Micromegas
detectors: a full microbulk readout plane formerly used in the CAST
experiment and a classical Micromegas structure without mesh. On the
other hand, two other samples (\#29-30 of table~\ref{res}) were
screened corresponding just to the raw foils used in the fabrication
of microbulk readouts, consisting of kapton metalized with copper on
one or both sides. The raw materials (kapton and copper, mainly)
were confirmed to be very radiopure; the numbers for the treated
foils show similar limits or values just at the limit of the
sensitivity of the germanium measurement. Despite their importance,
these bounds were still relatively modest when expressed in
volumetric terms, due to the small mass of the samples. New activity
measurements for the microbulk Micromegas and Cu-kapton-Cu foil
samples (previously measured with Ge spectroscopy) were carried out
profiting the great capabilities of the BiPo-3 detector operating at
LSC (\#28,31 of table~\ref{res}). For both cases, only limits to the
contamination in $^{208}$Tl and $^{214}$Bi can be deduced, which
improve the Ge spectrometry limits by more than 2 orders of
magnitude, pointing to contaminations at the level of, or below,
$\sim$0.1~$\mu$Bq/cm$^{2}$ \cite{irastorza16dbd}. A more sensitive
measurement for microbulk Micromegas produced at CERN was made in
2016, using two capsules of the BiPo-3 detector
(30$\times$30~cm$^{2}$ each); results shown in \#33 of
table~\ref{res} point to a very significant reduction of the upper
limits of both $^{208}$Tl and $^{214}$Bi. All this confirms our
expectations that microbulk readouts contain radioactivity levels
well below typical components in very low background detectors.

As the BiPo-3 detector can only quantify activity from the U and Th
chains, but a non-negligible K content seemed to be present in the
first analysis of microbulk Micromegas~\cite{mmradiopurity}, a new,
more sensitive analysis of the $^{40}$K activity was undertaken
since its contribution in the background model was found to be
important \cite{iguaz16}. A sample of the same microbulk Micromegas
analyzed in the BiPo-3 detector was screened using a germanium detector
(\#32 of table~\ref{res}) deriving upper limits for all the common
radioisotopes. Limits set for $^{238}$U and $^{232}$Th chains are
higher than those derived from BiPo-3 measurement as expected. The
limit for $^{40}$K is $<$2.3~$\mu$Bq/cm$^{2}$, a factor 25 lower
than the estimated value in~\cite{mmradiopurity}; even if very
promising, it was not conclusive yet as the analyzed sample had not
the holes produced by the potassium compound which could be
responsible of a $^{40}$K contamination. Then, a new, more massive
sample of readouts from CERN with a total surface of 12372.75~cm$^{2}$ was analyzed. It consisted of faulty GEMs glued on kapton, produced as
microbulk Micromegas, which have gone through some chemical baths
involving potassium compounds to create the kapton holes. This sample has been screened up to three times:
\begin{itemize}
\item In the first screening in a germanium detector, as shown in \#34 of table~\ref{res}, upper limits have been set for
all the common radioisotopes except for $^{40}$K; those of
$^{232}$Th and $^{238}$U activity are, as expected, higher than the
ones obtained from the BiPo-3 detector (about a factor of 5.7 and
2, respectively). The signal from $^{40}$K is clear and its
activity has been quantified as (3.45$\pm$0.40)~$\mu$Bq/cm$^{2}$
(corresponding to a specific activity of (258$\pm$30)~mBq/kg). This
seemed to confirm that the potassium content is related to the
production of holes in the readout.
\item In an attempt to reduce potassium, the same sample with holes was cleaned in
water baths at CERN (dipped one week in tap water followed by a long rinse with deionized (DI) water) and a new screening to assess the effect of this
treatment using the same detector was made. As presented in \#35 of table~\ref{res}, the measured $^{40}$K
activity is (0.84$\pm$0.16)~$\mu$Bq/cm$^{2}$, reduced by a factor 4, but an important uranium activity is unexpectedly obtained. The ratio of activities of the mothers of the $^{238}$U and $^{235}$U chains is 20.6, in very good agreement with the expectation for natural uranium. The origin of the relevant uranium contamination found, at the level of (564$\pm$62)~mBq/kg of $^{238}$ or (45.8$\pm$5.0)~ppb of U, is unknown. A possibility is that it is related to the tap water used for baths\footnote{Levels of uranium in drinking water are of the order of 10~$\mu$g/l or 10~ppb, being possible a large dispersion around this value.}. Presence of $^{7}$Be (T$_{1/2}$=53.22~days, decaying by Electron Capture to ground or excited states) has been identified too thanks to the 477.6~keV gamma emission; due to its half-life, comparable to the measuring time, no direct estimate of the activity has been attempted.
\item A second cleaning was made in the same sample in order to assess its effect on both the uranium and the potassium content quantified. In this case the baths were performed only with DI water (for one week the sample was cleaned with DI water, changed each day and heated to 60$^{o}$C). The obtained results (\#36 of table~\ref{res}), with no significant change in the derived activities, point to a null effect of the new procedure for both the $^{40}$K and U activities.
\end{itemize}
Following these results, alternative cleaning procedures and even the possibility of etching kapton by plasma, totally avoiding potassium compounds, are being studied. It is also worth noting that no indication of a possible $^{210}$Pb contamination has been found in any of the measurements for this sample, as there is no excess over background at the 46.5~keV peak.

A number of other samples involved in various Micromegas fabrication
processes have also been measured \cite{irastorza16dbd}. A sample of
pyralux sheets, used in the construction of bulk Micromegas, showed
good radiopurity first in germanium screening (\#37 of
table~\ref{res}) and afterwards using the BiPo-3 detector (\#38 of
table~\ref{res}). This result is of interest for the development of
radiopure bulk Micromegas. A kapton-epoxy foil (AKAFLEX, from
Krempel GmbH) used in the microbulk fabrication process (to join
several kapton layers in more complex routing designs) has been
measured in BiPo-3 (\#40 of table~\ref{res}) showing similar values
to the previous samples. In addition, a sample of adhesive Isotac 3M
VHB used in Micromegas readouts has been screened with a germanium
detector (\#39 of table~\ref{res}) setting upper limits for all the
common radioisotopes.
A sample of the stainless steel mesh used in the bulk Micromegas
firstly used in TREX-DM (produced by Somacis) has been screened
(\#41 of table~\ref{res}) deriving only upper limits for all the
analyzed radioisotopes.

Alternative production procedures for Micromegas are being explored.
Two units of Micromegas (diameter 10~cm, thickness 500~$\mu$m, made
of Si covered by silicon oxide SU8 and aluminum) produced at the
Centro Nacional de Microelectronica, Barcelona, have been screened
(\#42 of table~\ref{res}). Upper limits have been set for all the
common radioisotopes, pointing to a radiopure production process at
CNM in this first approach.

\footnotesize
\onecolumn
\begin{landscape}
\begin{longtable}{p{0.2cm}p{4.3cm}p{1.6cm}p{1.2cm}p{1.7cm}p{1.5cm}p{1.5cm}p{1.5cm}p{1.3cm}p{1.7cm}p{0.9cm}p{1cm}}
 \hline
\textbf{\#} & \textbf{Material,Supplier} &  \textbf{Method} & \textbf{Unit} & \textbf{$^{238}$U} & \textbf{$^{226}$Ra} & \textbf{$^{232}$Th} &\textbf{$^{228}$Th} & \textbf{$^{235}$U}& \textbf{$^{40}$K}  & \textbf{$^{60}$Co}& \textbf{$^{137}$Cs}\\
 \hline
\endfirsthead
(Continuation)\\
\hline
\textbf{\#} & \textbf{Material,Supplier} &  \textbf{Method} & \textbf{Unit} & \textbf{$^{238}$U} & \textbf{$^{226}$Ra} & \textbf{$^{232}$Th} &\textbf{$^{228}$Th} & \textbf{$^{235}$U}& \textbf{$^{40}$K}  & \textbf{$^{60}$Co}& \textbf{$^{137}$Cs}\\
\hline
\endhead
&(Follows at next page)\\
\endfoot
\endlastfoot
1 & Pb, Mifer &  GDMS & mBq/kg & 0.33 &&0.10&&& 1.2&&\\
2 & OFE Cu, Luvata  & GDMS & mBq/kg& $<$0.012&& $<$0.0041&&& 0.061&&\\
3 & ETP Cu, Sanmetal &GDMS &mBq/kg & $<$0.062&& $<$0.020 &&&&&\\
4 & ETP Cu, Sanmetal & Ge Oroel & mBq/kg & $<$27 & $<$1.0 & $<$1.1 & $<$0.76 & $<$0.56 & $<$3.1 & 0.24$\pm$0.05 & $<$0.29 \\
5 & PFA tube, Emtecnik & Ge Paquito& mBq/m & $<$31 & $<$0.58 &  $<$0.53 & $<$0.34 & $<$0.29 & $<$2.6 & $<$0.16 & $<$0.18 \\
6 & PTFE tube, Tecnyfluor & Ge Paquito & mBq/m & $<$19 & $<$0.48 & $<$0.54 & $<$0.41 & $<$0.26 & $<$2.5 & $<$0.14 & $<$0.17 \\
\hline
7 & Kapton-Cu PCB, LabCircuits  & Ge Paquito & $\mu$Bq/cm$^{2}$ & $<$42 & $<$1.3 & $<$1.1 & $<$0.66 & $<$0.41 & $<$4.0 & $<$0.24& $<$0.28  \\
8 & Epoxy Hysol, Henkel  & Ge Paquito& mBq/kg & $<$273 & $<$16 &$<$20& $<$16 & & $<$83& $<$4.2& $<$4.5 \\
9 & SM5D resistor, Finechem    & Ge Paquito& mBq/pc& 0.4$\pm$0.2 & 0.022$\pm$0.007 & $<$0.023 & $<$0.016  & 0.012$\pm$0.005 & 0.17$\pm$0.07& $<$0.005 & $<$0.005 \\
10 & Mylar, Goodfellow & Ge Paquito&  $\mu$Bq/cm$^{2}$ & $<$29 & $<$0.59 & $<$0.80 & $<$0.36 & $<$0.29 & $<$3.3 & $<$0.18 & $<$0.21 \\
11 & Nylon (3D printer), CNM & Ge Latuca &  mBq/kg & $<$436 & $<$9.2 & $<$11 & $<$3.4 & $<$2.6 & $<$29 & $<$1.0 & $<$1.2 \\
12 & Nylon (3D printer), CNM & ICPMS & mBq/kg & 36 & & 2.9 & & &&& \\
13 & Teflon, Sanmetal & ICPMS & mBq/kg &  $<$0.062 & &  $<$0.041 & & &&& \\
14 & Extruded PTFE, Gore & ICPMS & mBq/kg & $<$0.124 & &  $<$0.041 &&&&& \\
\hline
15 & Gold connectors, Fujipoly  & Ge Paquito& mBq/pc & $<$25 & 4.45$\pm$0.65 & 1.15$\pm$0.35 & 0.80$\pm$0.19 & & 7.3$\pm$2.6 & $<$0.1 & $<$0.4\\
16 & Silver connectors, Fujipoly  & Ge Paquito& mBq/pc & $<$55 & 5.68$\pm$0.81 & 6.1$\pm$1.1 & 6.17$\pm$0.72 & & 12.2$\pm$3.8 & $<$0.3 & $<$0.3\\
17  & Carbon connectors, Fujipoly  & Ge Paquito& mBq/pc & 14.5$\pm$6.0 & 2.77$\pm$0.38 & 1.17$\pm$0.23 & 1.14$\pm$0.14 & & 7.5$\pm$2.3 & $<$0.1 & $<$0.1\\
18 & Final Gold connectors, Fujipoly & Ge Paquito& mBq/pc& $<$12 & 2.80$\pm$0.38 & 0.49$\pm$0.10 & 0.58$\pm$0.09 & & 5.3$\pm$1.6 & $<$0.08 & $<$0.07 \\
19 & Kapton connectors, Samtec & Ge Paquito & mBq/pc & $<$3.6 & $<$0.065 & $<$0.072 & $<$0.040 & 0.043$\pm$0.015 & $<$0.32 & $<$0.020 & $<$0.021 \\
20 & Flat cable, Somacis  & Ge Paquito& mBq/pc & $<$14 & 0.44$\pm$0.12 & $<$0.33 & $<$0.19 & $<$0.19 & 1.8$\pm$0.7 & $<$0.09 & $<$0.10 \\
21 & Teflon cable, Druflon  & Ge Paquito& mBq/kg & $<$104 & $<$2.2 & $<$3.7 & $<$ 1.7 & $<$1.4 & 21.6$\pm$7.4 & $<$0.7 & $<$0.8 \\
22 & Coaxial cable, Axon & Ge Paquito& mBq/kg & $<$650 & $<$24 & $<$15 & $<$9.9 & $<$7.9 & 163$\pm$55 & $<$4.3 & $<$5.1 \\
23 & Electronic board, CEA & Ge Paquito& Bq/kg & 94$\pm$38 & 41.4$\pm$5.6 & 59$\pm$10 & 53.6$\pm$7.4 & & 19.5$\pm$6.1 & $<$0.67 & $<$1.1 \\
24 & AGET chips, CEA & Ge Paquito& mBq/pc & $<$8.7 & 0.48$\pm$0.07 & 0.16$\pm$0.06 & 0.47$\pm$0.09 & & 0.83$\pm$0.29 & $<$0.04 & $<$0.04 \\
25 & Ceramic AGET chips, CEA & Ge Paquito&  mBq/unit & (0.64$\pm$0.24)10$^{3}$ & 539$\pm$94 & 116$\pm$20 & 113$\pm$21 & & 43$\pm$14 & $<$2.2 &\\
 \hline
26 & Classical Micromegas, CAST & Ge Paquito& $\mu$Bq/cm$^{2}$ & $<$40 &&4.6$\pm$1.6&& $<$6.2& $<$46 & $<$3.1& \\
27 & Microbulk MM, CAST  & Ge Paquito& $\mu$Bq/cm$^{2}$ & 26$\pm$14&& $<$9.3&& $<$14 &57$\pm$25& $<$3.1& \\
28 & Microbulk MM, CAST & BiPo-3 & $\mu$Bq/cm$^{2}$ & & $<$0.134 & & $<$0.097 & & &  & \\
29 & Kapton-Cu foil, CERN  & Ge Paquito& $\mu$Bq/cm$^{2}$ & $<$11&& $<$4.6 &&$<$3.1& $<$7.7 &$<$1.6&\\
30 & Cu-kapton-Cu foil, CERN  & Ge Paquito& $\mu$Bq/cm$^{2}$ & $<$11 && $<$4.6&& $<$3.1& $<$7.7& $<$1.6& \\
31 & Cu-kapton-Cu foil, CERN & BiPo-3 & $\mu$Bq/cm$^{2}$& & $<$0.141 & & $<$0.033 & & &  & \\
32 & Microbulk MM, CERN & Ge Latuca & $\mu$Bq/cm$^{2}$ & $<$49 & $<$0.70 & $<$1.2 & $<$0.35 & $<$0.22 & $<$2.3 & $<$0.14 & $<$0.13 \\
33 & Microbulk MM, CERN & BiPo-3 & $\mu$Bq/cm$^{2}$ & & $<$0.045 & & $<$0.039 & &  & & \\
34 & Micromegas GEM, CERN & Ge Oroel & $\mu$Bq/cm$^{2}$ & $<$5.2 & $<$0.10 & $<$0.22 & $<$0.08 & $<$0.03 & 3.45$\pm$0.40 & $<$0.02 & $<$0.02\\
35 & Micromegas GEM 1$^{st}$ cleaning & Ge Oroel & $\mu$Bq/cm$^{2}$ & 7.41$\pm$0.81 & $<$0.21 & 0.19$\pm$0.05 & $<$0.11 & 0.36$\pm$0.04 & 0.84$\pm$0.16 & $<$0.02 & $<$0.03 \\
36 & Micromegas GEM 2$^{nd}$ cleaning & Ge Oroel & $\mu$Bq/cm$^{2}$ & 7.87$\pm$0.89 & $<$0.17 & 0.14$\pm$0.04 & 0.07$\pm$0.02 & 0.36$\pm$0.04 & 0.81$\pm$0.15 & $<$0.03 & $<$0.02 \\
37 & Pyralux, Saclay  & Ge Paquito& $\mu$Bq/cm$^{2}$& $<$19&$<$0.61&$<$0.63& $<$0.72& $<$0.19& 4.6$\pm$1.9& $<$0.10 &$<$0.14 \\
38 & Pyralux, Saclay & BiPo-3 & $\mu$Bq/cm$^{2}$& & $<$0.032 & & $<$0.036 & & & & \\
39 & Isotac adhesive, 3M & Ge Paquito& $\mu$Bq/cm$^{2}$ & $<$18 & $<$0.45 & $<$0.43 & $<$0.22 & $<$0.18 & $<$2.3 & $<$0.10 & $<$0.14  \\
40 & Kapton-epoxy foil, CERN & BiPo-3 & $\mu$Bq/cm$^{2}$& & $<$0.033 & & $<$0.022 & & & & \\
41 & Stainless steel mesh & Ge Paquito & $\mu$Bq/cm$^{2}$ &  $<$53 & $<$1.5  & $<$1.7 & $<$0.9 &  $<$0.6 & $<$8.7  & $<$0.3 & $<$0.5 \\
42 & Micromegas, CNM & Ge Paquito&  $\mu$Bq/cm$^{2}$ & $<$462 & $<$10 & $<$11 & $<$6.3 & $<$4.5 & $<$61 & $<$3.8 & $<$3.7 \\
\hline \caption{Activities measured for samples analyzed in the
radiopurity assessment program carried out for TREX-DM. Values from Ge detectors
reported for $^{238}$U and $^{232}$Th correspond to the upper part
of the chains and those of $^{226}$Ra and $^{228}$Th give activities
of the lower parts; reported errors correspond to $1\sigma$
uncertainties and upper limits are evaluated at 95\% C.L. Values from BiPo-3 detector come from the deduced activities of $^{214}$Bi and $^{208}$Tl isotopes and limits correspond to 90\% C.L.} \label{res}
\end{longtable}
\end{landscape}
\normalsize
\twocolumn

\section{Background contributions}
\label{bac}

\begin{figure}
\centering
\includegraphics[width=0.5\textwidth]{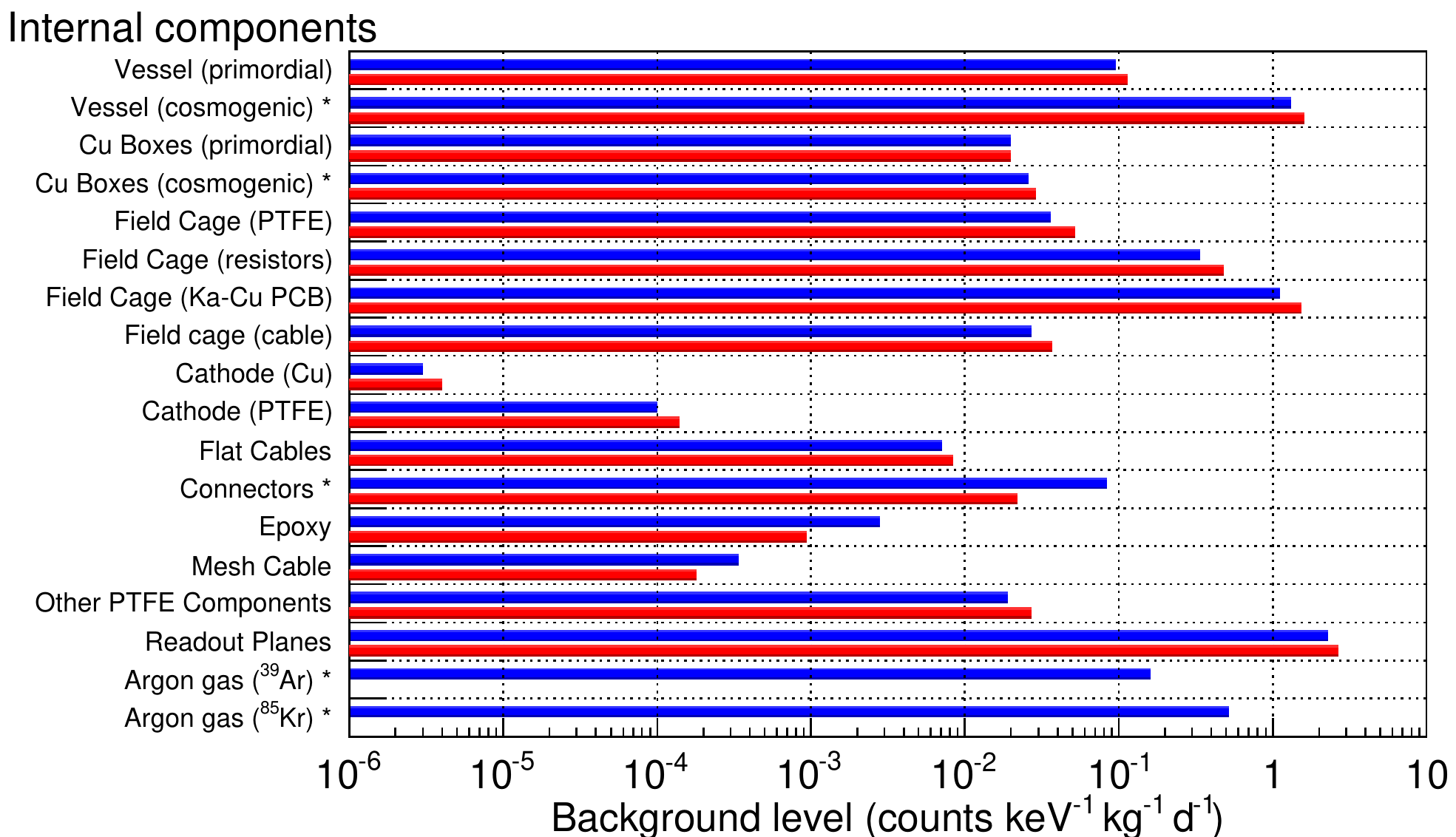}\\
\includegraphics[width=0.5\textwidth]{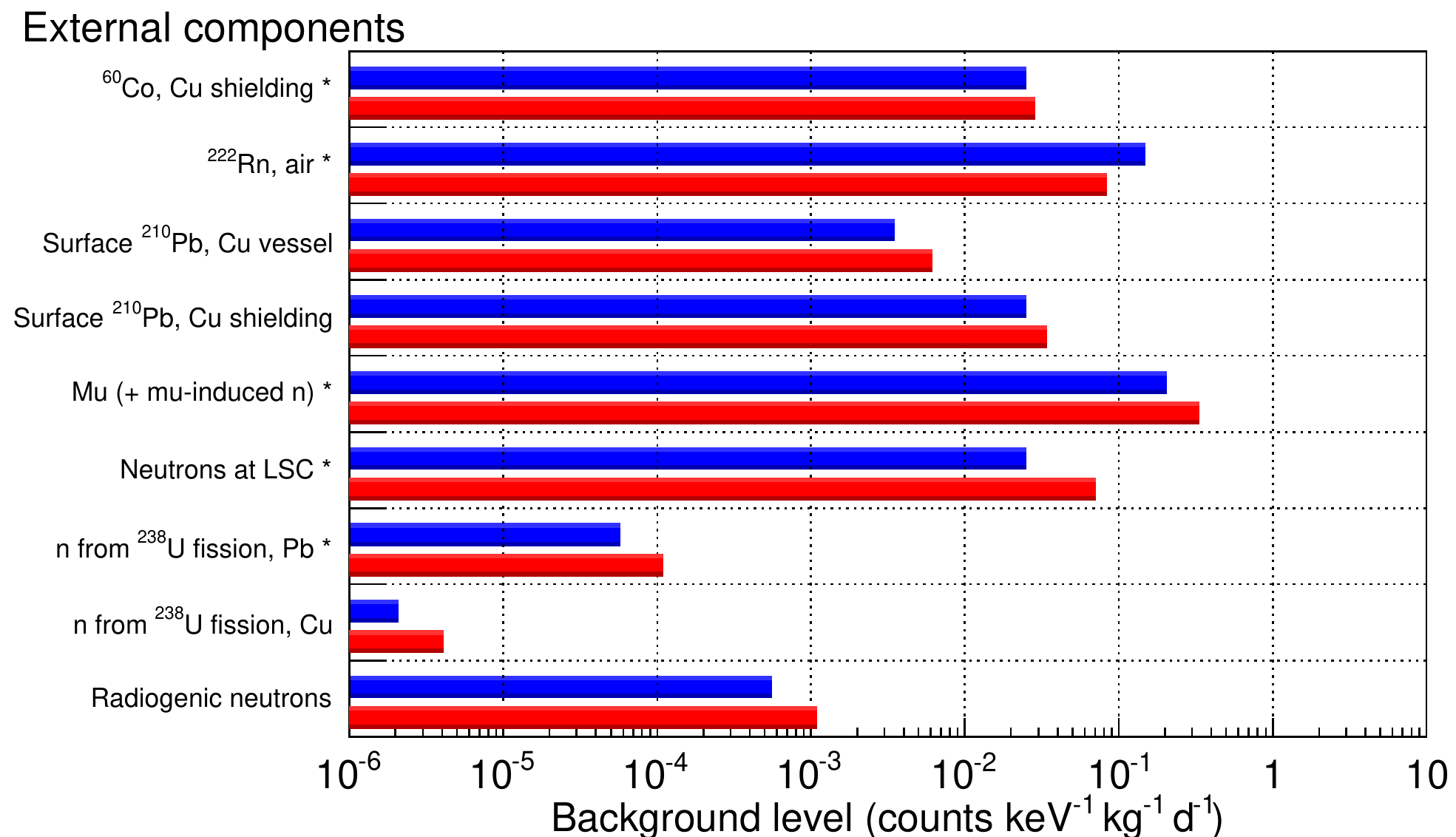}\\
\caption{Background rates (in counts keV$^{-1}$ kg$^{-1}$ d$^{-1}$) expected in the RoI (0.2-7~keV$_{ee}$) using Ar (blue bars) or Ne (red bars) mixtures in TREX-DM, as shown in table~\ref{rates}. Rates from contributions inside (top) and outside (bottom) the vessel are depicted. Contributions estimated using quantified background sources (not upper limits) are marked with *.} \label{figrates}
\end{figure}

\begin{figure}
\centering
\includegraphics[width=0.5\textwidth]{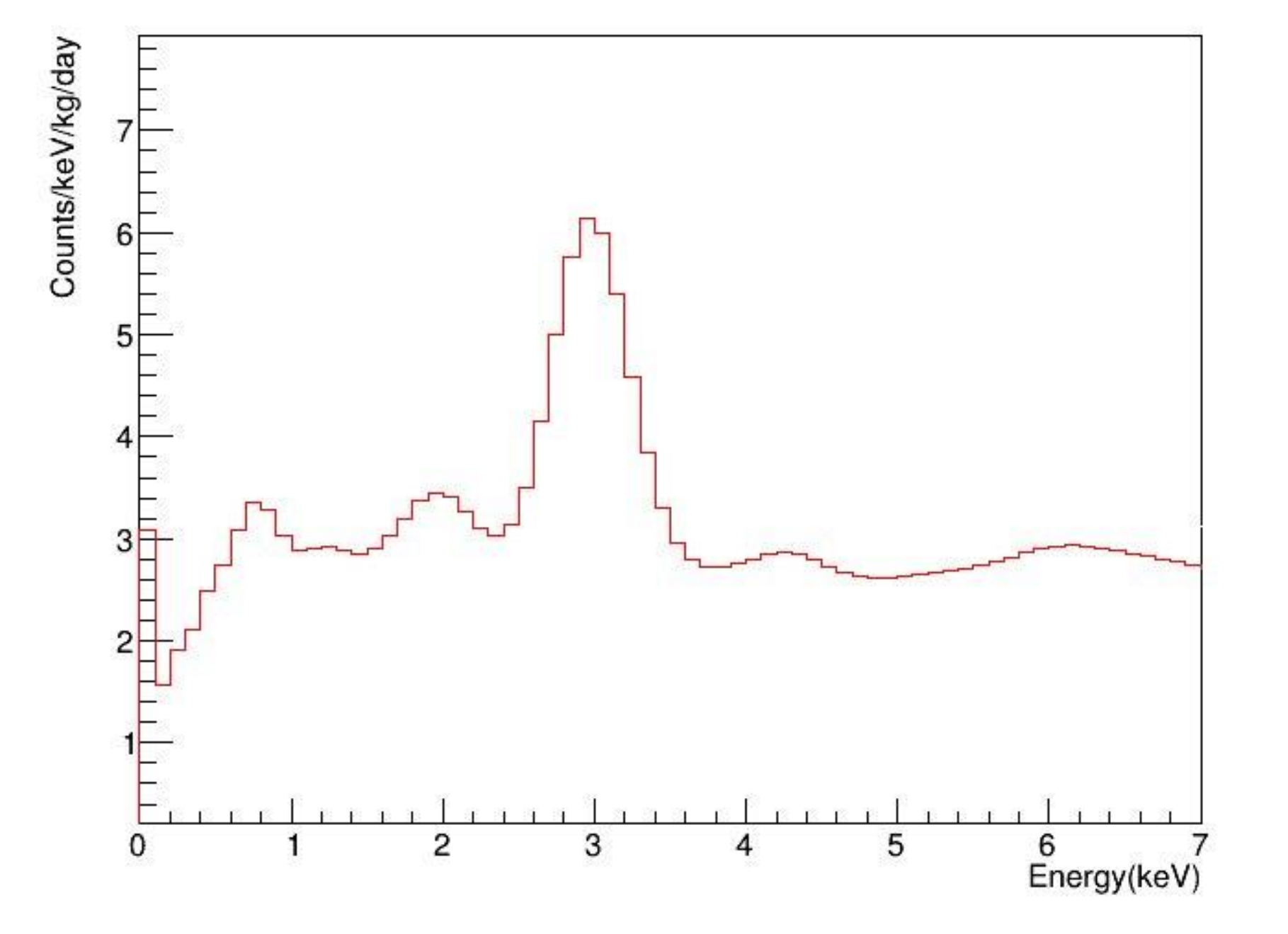}\\
\includegraphics[width=0.47\textwidth]{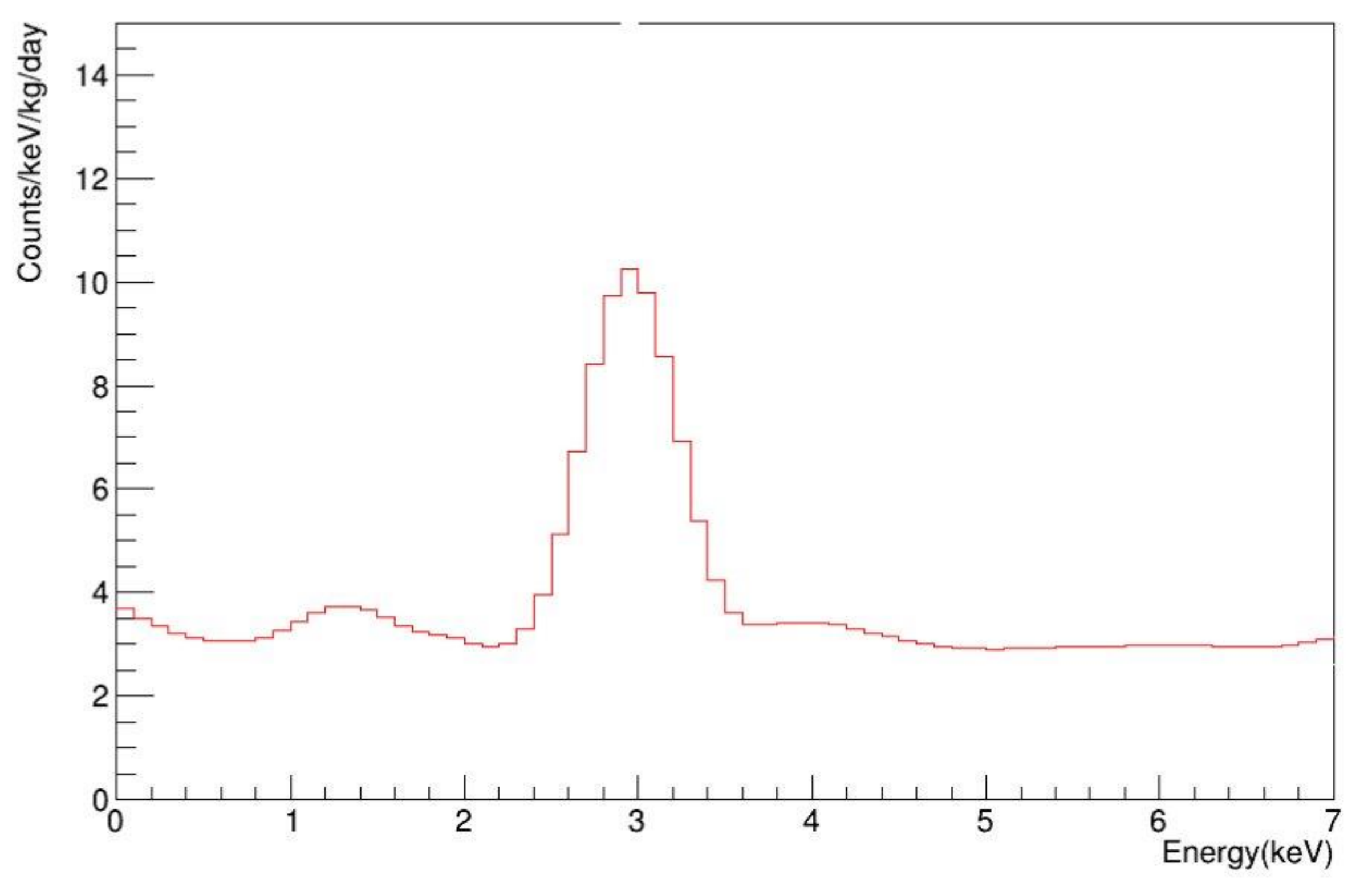}\\
\caption{Background spectra in the low energy region for Ar (top) and Ne (bottom) mixtures in TREX-DM, simulated for the dominant quantified internal activities: $^{40}$K from the readout planes and cosmogenic $^{60}$Co from the copper vessel. The experimental energy resolution of the detector has been considered to depict the spectra.} \label{figspectra}
\end{figure}

The main background sources have been simulated to evaluate their
contribution to the counting rate in the region of interest for dark
matter searches: radioactive isotopes in the elements of the set-up,
either primordial or cosmogenically produced; radon-induced
activity; and backgrounds at the LSC including gamma-rays, muons and
neutrons. Concerning the measured activities, whenever different values were obtained for the upper and lower parts of the $^{238}$U and $^{232}$Th chains (see table~\ref{res}), they were properly considered in the corresponding simulations. The background levels quoted in the following are referred
to a Region of Interest (RoI) of 0.2-7~keV$_{ee}$; the upper value is fixed to avoid the interference from the copper fluorescence peak at 8.0~keV while the low energy threshold assumed is equivalent
to 1~keV for Ar and Ne nuclear recoils. As made in \cite{iguaz16}, since the quenching factor of neither gaseous argon nor neon
has been measured yet, its parametrization in terms of the atom number and mass number from \cite{smithlewin} has been considered here. Results for all the components included in the
model and the different backgrounds at the laboratory are discussed in the rest of this section; the rates are summarized
in table~\ref{rates} and figure~\ref{figrates} and the energy spectra in the low energy region obtained for the dominant quantified  internal activities are shown in figure~\ref{figspectra}.

\subsection{Intrinsic radioactivity}

The simulation of the common radioactive isotopes, $^{40}$K and
those in the $^{238}$U and $^{232}$Th chains, from all the main
internal components (inside or close to the vessel) has been scaled by the measured activities. Table ~\ref{ratescomp} presents the individual contribution of each of the isotopes in the case of the considered Ne mixture (analogous results have been obtained for the Ar mixture), to identify the most relevant contributor for each simulated component of the set-up. As shown in table~\ref{rates}, most of the screened components finally
used in the set-up have been considered; it is worth noting that
many of the reported values are indeed upper limits to the estimated
background rate as only upper limits have been derived for the
activity of the corresponding element. The thorough selection
process of components and materials (described in section~\ref{rad})
has allowed to reduce to non-relevant levels the contributions of,
for instance, PTFE components, field cage resistors or silicone
connectors.

For the Micromegas readout planes, the measured value for $^{40}$K
activity before any special cleaning (as in \#31 of table~\ref{res}) has been taken into account. It gives a significant rate
and therefore the studies underway to reduce this activity, probably related to the process of creating
holes, are very important.


\subsection{Cosmogenic activity}

Cosmogenic activation of the set-up materials must be taken into
account too. Especially, that of the copper vessel, due to its large
mass, and the activation of the gas medium itself.

\subsubsection{Copper}
\label{baccoscop}
Production rates of cosmogenic isotopes in copper have been measured
\cite{Heusser:2006,Baudis:2015} and evaluated using different codes
\cite{Mei:2009,Cebrian:2010,activia,Zhang:2016}; although there is a
non-negligible dispersion in results, yields of tens of nuclei per
kg and day are expected for cobalt isotopes having long half-lives
(like 271.8~d for $^{57}$Co, 70.85~d for $^{58}$Co and 5.271~y for
$^{60}$Co). A simulation of the long-lived $^{60}$Co emissions from
the vessel and copper shields has been carried out. As discussed in
section~\ref{radves}, the cosmogenic activation of the TREX-DM
vessel (having a mass of $\sim$0.6~tons) has been quantified thanks
to the screening of a copper sample having the same exposure
history; this measured $^{60}$Co activity has been considered to
evaluate its contribution in the model, which is very relevant (see
table~\ref{rates}). The construction of a new vessel will allow to
significantly suppress this contribution. For the inner copper
shielding (with a mass of $\sim$2~tons) and the copper boxes
shielding the connectors (42~kg), the activity corresponding to
3~months of exposure to cosmic rays at sea level and the saturation
activity of (1.000$\pm$0.090)~mBq/kg deduced in~\cite{Heusser:2006}
has been considered; this contribution is not relevant, as shown in
table~\ref{rates}, thanks not only to the reduced cosmogenic activity but also to the shielding effect provided by the copper vessel for these emissions.

\subsubsection{Gas}
\label{secgas}

For the specific case of argon-based mixtures, the effect of
$^{39}$Ar, which decays by beta-emission (Q$=$565~keV) and has a
long half-life (239~y), has been evaluated. It is produced at
surface level by cosmogenic activation and the best way to avoid it
is extracting argon for underground sources. The lowest activities
have been obtained by the DarkSide collaboration using this
technique~\cite{argon}; assuming the measured activity from the DarkSide-50 data of 0.73~mBq/kg, the contribution to
TREX-DM background has been quantified (see table~\ref{rates}). This value of $^{39}$Ar activity means a reduction factor of about 1400 in comparison to the activity in atmospheric argon, which is at the level of 1~Bq/kg. Therefore, following the estimated contribution in table~\ref{rates} for $^{39}$Ar, it is straightforward to conclude that for operation with atmospheric argon the background level of TREX-DM would be totally dominated by this isotope, giving a rate of 219~counts keV$^{-1}$ kg$^{-1}$ d$^{-1}$ in the RoI. The contribution of $^{85}$Kr, decaying also by beta-emission (Q$=$687~keV) with a half-life of 10.76~y, has been estimated too, considering the measured activity by DarkSide~\cite{argon}, even if cryogenic distillation is expected to help to remove effectively $^{85}$Kr from argon.

Among the different radioactive isotopes that can be induced,
tritium in the detector medium could be a very relevant background
source for a dark matter experiment due to its decay properties: it
is a pure beta emitter with Q$=$18.591~keV and a long half-life of
12.312~y. Following the shape of the beta spectrum for the
super-allowed transition of $^{3}$H, 57\% of the emitted electrons
are in the range from 1 to 7~keV; these electrons are fully absorbed
in most of the, typically large, dark matter detectors. Some recent
studies on tritium production in materials of interest for dark
matter experiments can be found in
\cite{Mei:2009,edelweisscos,Zhang:2016}; there are some estimates in
argon but no information for neon, and therefore a calculation of
production rates in these two targets (assuming their natural
isotopic abundances) has been attempted, as presented in
\cite{tritium}.

\begin{figure}
  \includegraphics[width=.5\textwidth]{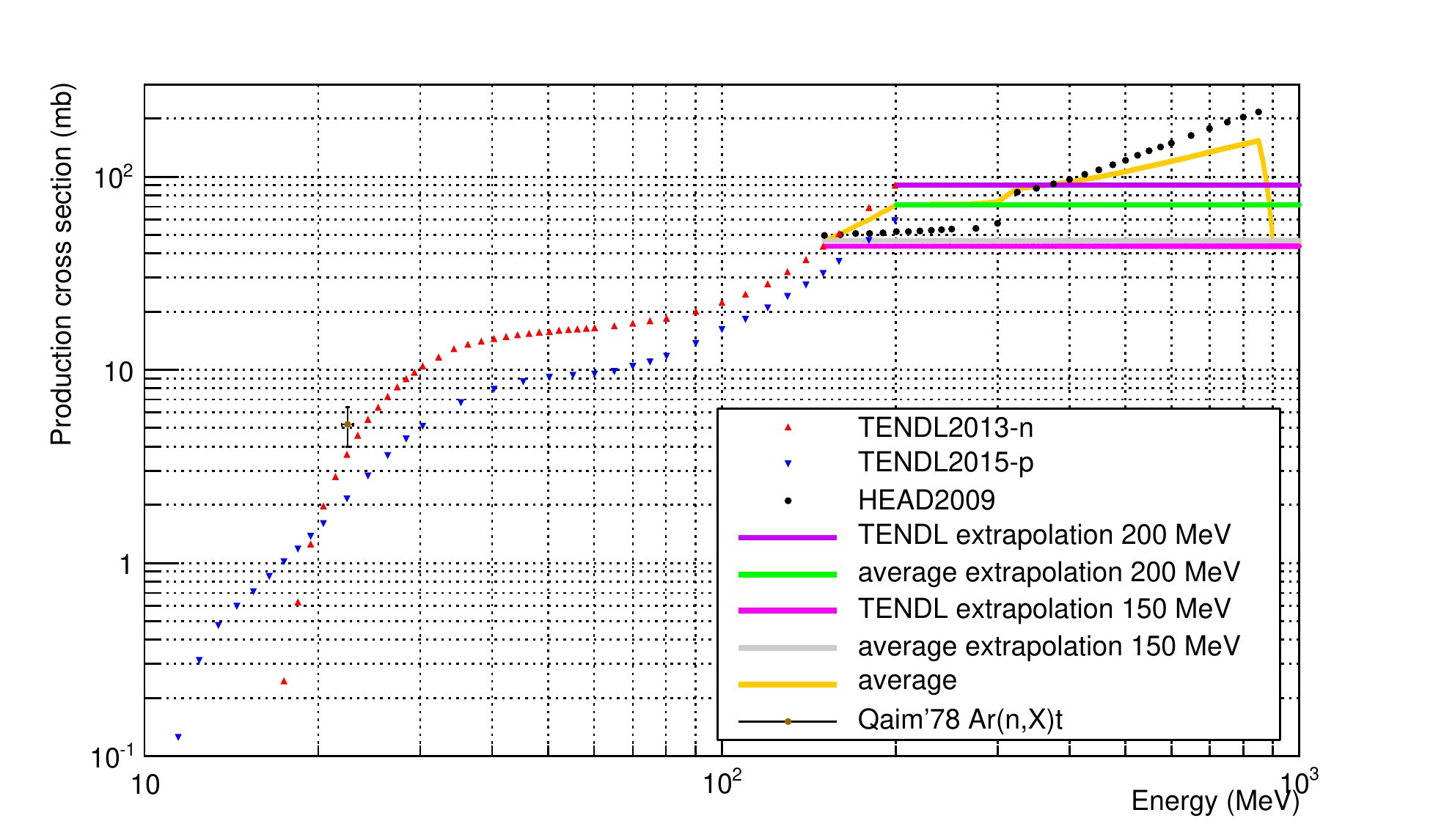}
  \includegraphics[width=.5\textwidth]{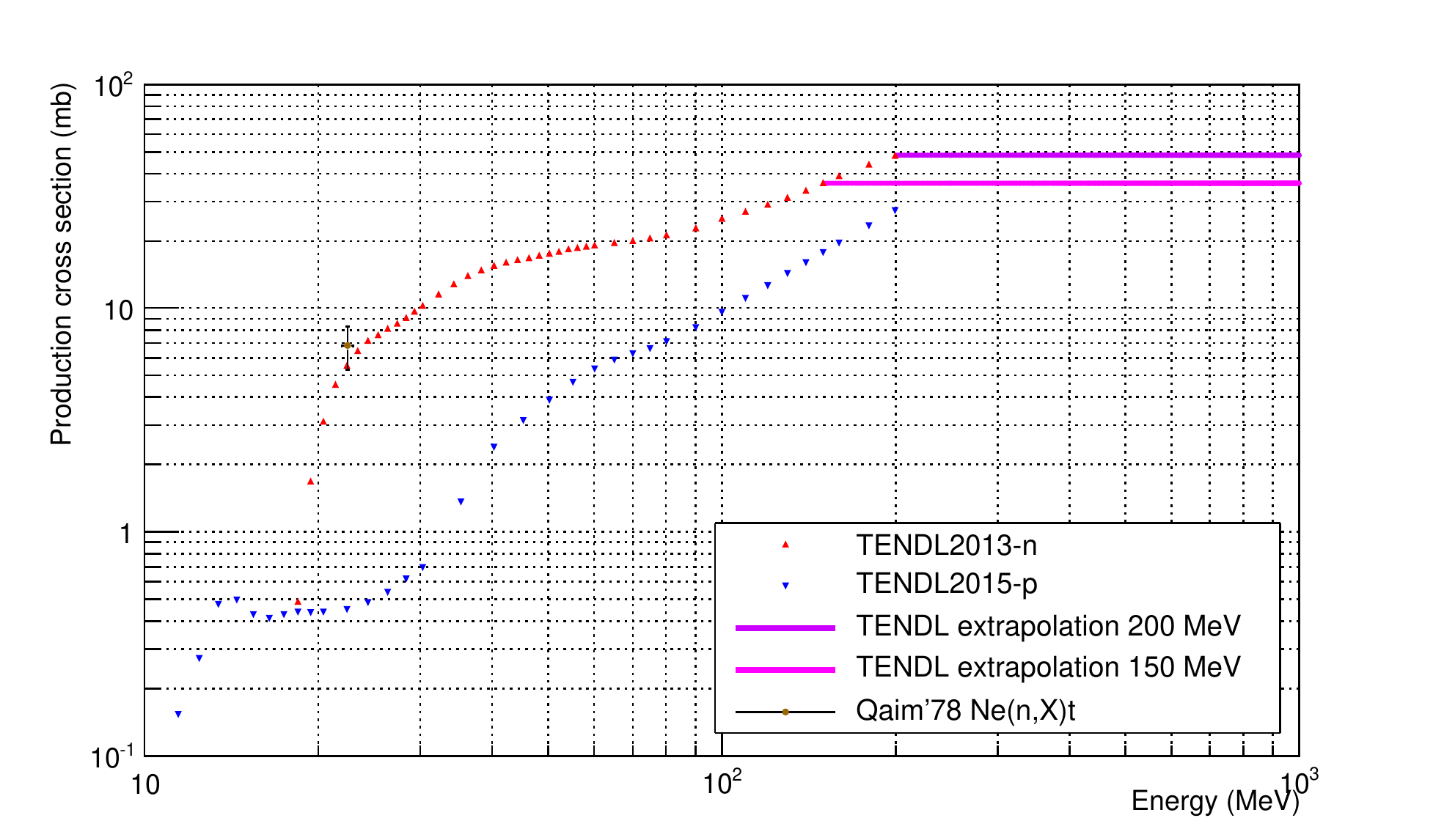}
  \caption{Comparison of excitation functions for the production of $^{3}$H  by nucleons for Ar (top)
  and Ne (bottom) taken from different sources: TENDL-2013 and HEAD-2009 libraries together
with several extrapolations considered at high energies; the
extrapolation of the average between the highest energy
cross-section from TENDL and the corresponding results from
HEAD-2009 is labeled as ``average extrapolation'' and the average at
each energy between TENDL data (and its extrapolation) and HEAD-2009
cross sections as ``average''. }
  \label{efArNe}
\end{figure}

The available information on the excitation function by nucleons has
been firstly collected, as shown in figure~\ref{efArNe}: only one
experimental result was found in the EXFOR database \cite{exfor}
from \cite{qaim} and cross sections were taken from the
TENDL-2013/2015 (TALYS-based Evaluated Nuclear Data Library) library
\cite{tendl} up to 150/ 200~MeV and from the HEAD-2009 (High Energy
Activation Data) library \cite{head} for higher energies up to
1000~MeV. Above 1~GeV, a constant production cross-section from the
last available energy has been assumed. For Ar, the excitation
functions from the two different libraries used at low and high
energies match reasonably well. Since the HEAD-2009 library does not
provide results for Ne, the last available cross-section value from
TENDL-2013 has been assumed for all the higher energies. Then, the
production rate was computed convoluting a selected excitation
function with the energy spectrum of cosmic neutrons at sea level,
using the parametrization from \cite{gordon}, following the
different approaches plotted in figure~\ref{efArNe} \cite{tritium}.
The maximum and minimum rates obtained in these approaches define an
interval, whose central value and half width have been considered as
the final results and their uncertainties for the evaluation of the
production rates of tritium in each target. Table~\ref{ratestritium}
summarizes the production rates obtained for Ar and Ne and presents
other results from the literature: the estimates in \cite{Mei:2009}
using TALYS 1.0 code and those in \cite{Zhang:2016} based on GEANT4
simulation and ACTIVIA \cite{activia}. The rate in Ar only from
TENDL-2013 library below 150 MeV is 47.7~kg$^{-1}$ d$^{-1}$, which
is in very good agreement with the value obtained in
\cite{Mei:2009}, since the library is also based on TALYS code. It is worth noting that the applied procedure to estimate tritium production rates in Ar and Ne gives a very
good agreement with the measured rates by EDELWEISS and CDMSlite experiments when applied to natural Ge \cite{tritium}.

\begin{table}
\begin{tabular}{lcccc}
\hline Target  & TENDL-2013 & TALYS  & GEANT4 & ACTIVIA   \\
& +HEAD-2009 & & & \\
& \cite{tritium} & \cite{Mei:2009}&  \cite{Zhang:2016} &  \cite{Zhang:2016} \\
\hline

Ar & 146$\pm$31& 44.4 & 84.9 & 82.9 \\
Ne & 228$\pm$16 & & &  \\ \hline
\end{tabular}
\caption{Production rates (in kg$^{-1}$d$^{-1}$) of $^{3}$H at sea
level calculated for Ar and Ne from TENDL-2013 and HEAD-2009
libraries as in \cite{tritium}; other available values taken from
the literature are shown for comparison.} \label{ratestritium}
\end{table}

Tritium emissions are fully absorbed in the gas producing a signal
indistinguishable from that of a dark matter interaction. If
saturation activity was reached for tritium, according to production
rates in table~\ref{ratestritium}, it would dominate the expected
background model with a contribution of 15 and 22~counts
keV$^{-1}$~kg$^{-1}$~day$^{-1}$ for Ar and Ne, respectively, in the
RoI. However, tritium is expected to be suppressed by purification
of gas and minimizing exposure to cosmic rays of the purified gas
should avoid any problematic tritium activation. The first experimental data in TREX-DM
will be extremely useful to confirm that tritium production is not a
relevant background source for the experiment.

On the other hand, in TREX-DM, mixtures of Ar or Ne with
1-2\%iC$_{4}$H$_{10}$ at 10~b are foreseen; tritium could be not
only cosmogenically induced in the noble gas but also be present in
the isobutane. No specific information about tritium content in
isobutane has been found. Assuming concentration as in
water\footnote{For natural surface waters there are about one
tritium atom per 10$^{18}$ atoms of hydrogen, following Ref.
\cite{sourcesoftritium}. The measured tritium activity is some
waters and the limits for drinking water give indeed higher
concentrations.}, this would also give a very relevant contribution
in the RoI of TREX-DM of 22 and 84~counts keV$^{-1}$~kg$^{-1}$~day$^{-1}$ for the considered Ar and Ne
mixtures, respectively. In any case, the obtention of isobutane from
underground gas sources, shielded from cosmic rays, avoids a
dangerous tritium content.

\subsection{Radon-induced activity}
\label{bacrad}

A simulation of the $^{222}$Rn in the air surrounding the vessel
inside the copper shielding has been performed. The measured
activity at hall A of LSC of (63$\pm$1)~Bq/m$^{3}$ \cite{bandac17}
has been considered; the implementation of a N$_{2}$ gas flux system
inside the shielding to avoid radon intrusion providing a factor
$\sim$100 reduction in the air $^{222}$Rn activity allows to reduce
this contribution to a non-relevant level (see table~\ref{rates}). Even if radon inside the shielding has not been directly quantified, the reduction factor assumed here and in other experiments operated in Canfranc with similar shielding and nitrogen flux is supported by the good agreement between models and measured data for peaks due to $^{222}$Rn emissions \cite{anais}. In any case, this assumption is not very relevant because if TREX-DM operates using the radon-free air facility available at the LSC a bigger radon suppression factor at the level of about 1000 would be at reach.

An estimate of the contribution to the TREX-DM background of a long
exposure to air-borne radon of components of the set-up has been
attempted, evaluating the effect of the creation of a long-lived
$^{210}$Pb contamination and its progeny. This study has been
carried out for the copper vessel, having a large surface
($\sim$1.4~m$^{2}$) exposed to a normal atmosphere for a long time,
and for the inner part of the copper shielding ($\sim$2.9~m$^{2}$).
For the induced $^{210}$Pb activity on the copper surface, the upper
limit set from the screening of a sample of the vessel material
described in section~\ref{radves} has been considered. It is worth
noting that the considered limit on $^{210}$Pb is $\sim$19 times
larger than the measured $^{210}$Po surface activity on an ETP
copper exposed to radon for years \cite{Zuzel2017} and $\sim$10-30
times larger than the calculated activity assuming 2~y of exposure
to an activity of 15~Bq/m$^{3}$ of $^{222}$Rn following the results
in \cite{Stein2017},\cite{Clemenza:2011}; therefore, it is
considered a very conservative value. Emissions of $^{210}$Pb from
the copper surfaces have been simulated and, following results in
table~\ref{rates}, the contribution is not worrisome.

\subsection{Gamma background}
The environmental gamma flux contribution from $^{238}$U, $^{232}$Th
and $^{40}$K emissions has been evaluated by simulating the
corresponding photons through the whole TREX-DM set-up and scaling
by the measured flux in hall A of LSC, as reported in
\cite{nextsensitivity}. Due to lack of statistics, modest upper limits for the counting rate in the RoI (0.2-7~keV$_{ee}$) at the level of 10-20~counts keV$^{-1}$
kg$^{-1}$ d$^{-1}$) have been set from this simulation. But the 20-cm-thick lead shielding together with the additional 11~cm copper layer from vessel and inner shielding allows for a reduction of several orders of magnitude of the environmental gamma flux and this contribution can be safely neglected.

\subsection{Muons}
\label{secmuo}

The muon contribution has been simulated assuming the measured muon
flux at LSC of $\sim4\times10^{-3}$~s$^{-1}$ m$^{-2}$, as
in~\cite{Luzon:2006sh}. The muon energy spectrum evaluated for the
Canfranc depth (with mean energy of 216~GeV) and the corresponding
angular distribution have been considered following~\cite{grieder}.
Muons are launched from a virtual wall through the whole TREX-DM
set-up. The counting rate derived in the RoI without applying any
discrimination method is high, but following the previous results
derived for TREX-DM by analyzing the signal topology
in~\cite{iguaz16}, only 5.4 (3.4)\% of events survive the cuts for
Argon (Neon); this makes the muon contribution non-dominant (see
table~\ref{rates}), even without the implementation of a muon veto.
It must be noted that the production of neutrons by muons is
implemented in the simulation and therefore the contribution of
these high energy neutrons produced in the lead shielding is
included in the background level reported for muons.

\subsection{Neutrons}
\label{secneu}

The effect of neutrons from other origins has been analyzed for
TREX-DM too. First, environmental neutrons at LSC have been
simulated from a virtual sphere containing the set-up with the
40-cm-thick moderator and the results have been normalized to the
neutron flux measured at LSC in~\cite{Jordan:2013} (similar to that
presented in \cite{Carmona:2004qk}); the evaporation spectrum has
been considered for the neutron energy sampling. A preliminary
analysis of the tracks left in the chamber points to a rejection
factor of $\sim$10\% of the events in the RoI. The obtained rates
are shown in table~\ref{rates}; the neutron moderator makes the
contribution from these neutrons irrelevant. Concerning
muon-induced neutrons in the rock surrounding the laboratory, the
expected flux is $\sim$3 orders of magnitude lower than that of the
environmental neutrons~\cite{Carmona:2004qk}.

Radiogenic neutrons produced by ($\alpha$,n) reactions due to
$^{238}$U and $^{232}$Th primordial impurities (or spontaneous
fission of the former) in the set-up materials are other of the
neutron sources of background in WIMP direct detection experiments.
In particular, neutrons having also an evaporation spectrum have
been simulated from the lead shielding and the copper vessel to
evaluate the contribution from the fission of their $^{238}$U
contaminations. The information on this isotope activity for copper
and lead presented in table~\ref{res} has been considered together
with the values of (2.4$\pm$0.2)~neutrons per fission and
(5.45$\pm$0.04)$\times$10$^{-7}$ fissions per decay \cite{nfis}. The
deduced counting rates, shown in table~\ref{rates}, are negligible.

Additionally, an estimate of the contribution of these radiogenic
neutrons in other materials of the TREX-DM set-up has been attempted
using a simulation of neutrons having a fission energy spectrum and calculating the expected neutron rate from neutron
yields estimated in previous works for the materials. In particular,
neutron yields for copper, steel, teflon and polyethylene from
\cite{tomasello} have been used; yields for Ar and Ne are available in \cite{mei}.
Results in \cite{tomasello} were obtained using the SOURCES4A code
with ($\alpha$,n) reaction cross-sections calculated using
EMPIRE2.19, while calculations in \cite{mei} are based on
cross-sections obtained with the TALYS code. Secular equilibrium in
the natural chains is assumed in both references. Yields in
\cite{tomasello} include also spontaneous fission of $^{238}$U. In
\cite{mei} it is reported that there is no ($\alpha$,n) neutron
yield in lead due to a very high Coulomb barrier. A code
(NeuCBOT) for computation of neutron yields and the corresponding
estimates for different materials have been presented too in
\cite{westerdale}; it is based on TALYS for cross sections and uses
ENSDF for alpha decay data and SRIM code for considering the
stopping power of alpha particles. It is reported that NeuCBOT tends
to predict yields systematically higher by $\sim$30\% than other
calculations based on SOURCES4A or available measured yields.
Table~\ref{tabn} summarizes the considered neutron yields and the
corresponding estimated counting rates in TREX-DM. The whole mass of
each material in the set-up has been considered: copper in vessel,
small components and shielding structure, teflon from field cage and
other components, steel in the shielding structure and polyethylene
used as neutron moderator. The following activities for $^{238}$U
and $^{232}$Th (using the highest value of those corresponding to
the different isotopes in the chain, if available) have been
assumed: the deduced values for TREX-DM for
copper and teflon (\#2 and \#13 in table~\ref{res}), measured values for S275 steel (as used in
TREX-DM shielding structure) by NEXT \cite{lrt2015} and upper limits
derived by ICPMS for both U and Th in polyethylene (from a different
supplier) by NEXT \cite{next4}. As shown in table~\ref{tabn}, neutrons from steel are the most relevant; the upper limit to the
total estimated contribution to the background rate is of the order
of 10$^{-3}$~counts~keV$^{-1}$ kg$^{-1}$~d$^{-1}$, and therefore, it can be concluded that radiogenic neutrons are not a dominant background source for
TREX-DM. This contribution is also shown in table~\ref{rates}. No reference of U and Th content in
the gases (Ar or Ne) is available; but it has been checked, using the yields in \cite{mei}, that concentrations of U and Th about 1~ppm would be required to produce rates at the level of 10$^{-3}$~counts~keV$^{-1}$ kg$^{-1}$~d$^{-1}$ comparable to those from the other materials.

The contribution from muons and environmental neutrons is under
control in the simulated conditions, following the rates in
table~\ref{rates}.  Although the precise estimates of some external
background sources are still underway, it seems that the designed
shielding is enough to reduce to non-relevant levels the
corresponding contribution.

\begin{landscape}
\begin{table}
\begin{center}
\begin{tabular}{lccc}
\hline Component  & Reference &  Argon &  Neon  \\
\hline
Vessel (primordial) & \#2 & $<$0.095  &  $<$0.114  \\
Vessel (cosmogenic) & \#4 & 1.31 & 1.60  \\
Copper Boxes (primordial) & \#2 & $<$0.020 & $<$0.020 \\
Copper Boxes (cosmogenic) & see Section~\ref{baccoscop} & 0.026 & 0.029 \\
Field Cage (PTFE) & \#13 &  $<$0.036 & $<$0.052  \\
Field Cage (resistors) & \#9 &  $<$0.34 &  $<$0.48 \\
Field Cage (kapton-Cu PCB)& \#7 &  $<$1.12 &  $<$1.54 \\
Field Cage (cable)  & \#20 &  $<$0.027 &  $<$0.037 \\
Cathode (copper) & \#2 &  $<$3$\times$10$^{-6}$  &  $<$4$\times$10$^{-6}$ \\
Cathode (PTFE) & \#13 &  $<$1.0$\times$10$^{-4}$  &  $<$1.4$\times$10$^{-4}$ \\
Flat Cables & \#19 &  $<$0.0071 &  $<$0.0084 \\
Connectors & \#18 &  0.083 &  0.022 \\
Epoxy & \#8 &  $<$0.0028 &  $<$0.00094 \\
Mesh Cable & \#21 &  $<$3.4$\times$10$^{-4}$ & $<$1.8$\times$10$^{-4}$ \\
Other PTFE Components & \#13 &  $<$0.019 &  $<$0.027 \\
Readout Planes & \#30,37&  $<$2.30 &  $<$2.68 \\
Argon gas ($^{39}$Ar) & \cite{argon} & 0.16 & \\
Argon gas ($^{85}$Kr) & \cite{argon} & 0.52 & \\
\hline
Total from internal components && $<$6.1 & $<$6.6 \\
Total from quantified internal activity && 4.25$\pm$0.36 & 4.20$\pm$0.43\\
\hline
Cosmogenic $^{60}$Co in Cu shielding & see Section~\ref{baccoscop} & 0.0250$\pm$0.0018 & 0.0288$\pm$0.0020 \\
$^{222}$Rn in air & \cite{bandac17} &0.1495$\pm$0.0024 & 0.0841$\pm$0.0013 \\
Surface $^{210}$Pb on Cu vessel & see Section~\ref{bacrad} & $<$3.5$\times$10$^{-3}$ &  $<$6.2$\times$10$^{-3}$  \\
Surface $^{210}$Pb on Cu shielding  & see Section~\ref{bacrad} & $<$0.025 & $<$0.034\\ 
Muons (+ muon-induced neutrons) & \cite{Luzon:2006sh} &0.205$\pm$0.021  & 0.336$\pm$0.034 \\
Neutrons at LSC & \cite{Jordan:2013} &(2.52$\pm$0.22)$\times$10$^{-2}$ & (7.06$\pm$0.61)$\times$10$^{-2}$ \\
Neutrons from $^{238}$U fission in Pb & see Section~\ref{secneu} &(5.82$\pm$0.39)$\times$10$^{-5}$ & (1.094$\pm$0.074)$\times$10$^{-4}$  \\
Neutrons from $^{238}$U fission in Cu & see Section~\ref{secneu} & $<$2.1$\times$10$^{-6}$  & $<$4.1$\times$10$^{-6}$  \\
Radiogenic neutrons in Cu, PTFE, steel and polyethylene & Table~\ref{tabn} &  $<$5.6 10$^{-4}$ & $<$1.1 10$^{-3}$ \\ \hline
Total from external components && $<$0.43 & $<$0.56  \\
Total from quantified sources && 0.40$\pm$0.02 & 0.52$\pm$0.03  \\
 \hline
\end{tabular}
\end{center}
\caption{\label{rates} Background rates (in counts keV$^{-1}$ kg$^{-1}$ d$^{-1}$) expected in the RoI (0.2-7~keV$_{ee}$) from activity in components and backgrounds at LSC using Ar or Ne mixtures in TREX-DM. The numbers with \# at the second column refer to table~\ref{res}, indicating the activity values considered. Two total rates are presented internal and external components: one calculated only from the quantified sources and another where all the contributions including upper limits have been taken into account.}
\end{table}
\end{landscape}

\begin{landscape}
\begin{table}
\begin{center}
\begin{tabular}{lccccc}
\hline Component  & Reference &  $^{238}$U &  $^{232}Th$ & $^{40}$K & $^{^{60}}$Co  \\
\hline
Vessel  & \#2,4 & $<$0.054 & $<$0.034 & 0.026 & 1.60 \\ 
Copper Boxes  & \#2, see Section~\ref{baccoscop} & $<$0.014 & $<$0.0064& 3.7$\times$10$^{-4}$ & 0.029 \\ 
Field Cage (PTFE) & \#13 & $<$0.026  & $<$0.026 & $<$7.3$\times$10$^{-5}$ &  \\ 
Field Cage (resistors) & \#9 & 0.20 & $<$0.22 & 0.053  & \\ 
Field Cage (kapton-Cu PCB)& \#7 & $<$0.82  & $<$0.64 & $<$0.087 & \\
Field Cage (cable)  & \#20 & $<$0.015 & $<$0.018 & 0.0050 & \\ 
Cathode (copper) & \#2 & $<$2.1$\times$10$^{-6}$ & $<$1.1$\times$10$^{-6}$ &  6.8$\times$10$^{-7}$ & \\  
Cathode (PTFE) & \#13 & $<$6.7$\times$10$^{-5}$  & $<$7.1$\times$10$^{-5}$ & $<$2.3$\times$10$^{-7}$  & \\ 
Flat Cables & \#19 & 0.0049 & $<$0.0015 & 0.0021 & \\
Connectors & \#18 & 0.0071  & 0.0026 & 0.013& \\ 
Epoxy & \#8 & $<$1.3$\times$10$^{-4}$   & $<$1.9$\times$10$^{-4}$  & $<$6.2$\times$10$^{-4}$  & \\ 
Mesh Cable & \#21 & $<$2.2$\times$10$^{-5}$ & $<$1.7$\times$10$^{-5}$ & 1.4$\times$10$^{-4}$& \\ 
Other PTFE Components & \#13 & $<$0.013   & $<$0.013 & $<$3.7$\times$10$^{-5}$ & \\ 
Readout Planes & \#30,37& $<$0.36  & $<$0.064 & 2.3 & \\ 
\hline
\end{tabular}
\end{center}
\caption{\label{ratescomp} Background rates (in counts keV$^{-1}$ kg$^{-1}$ d$^{-1}$) expected in the RoI (0.2-7~keV$_{ee}$) from the activity of each one of the considered contaminants in components using the Ne mixture in TREX-DM. The numbers with \# at the second column refer to table~\ref{res}, indicating the activity values considered. Analogous results have been obtained for the Ar mixture.}
\end{table}
\end{landscape}

\begin{landscape}
\begin{table}
\begin{tabular}{lcccccc}
\\
\hline &  Neutron yield \cite{tomasello} &   Mass &   Concentration &   Total yield & Rate in Ar & Rate in Ne \\
& (ppm$^{-1}$ g$^{-1}$ y$^{-1}$) & (kg) &(ppm)& (neutrons d$^{-1}$)& (counts~keV$^{-1}$ kg$^{-1}$~d$^{-1}$)& (counts~keV$^{-1}$ kg$^{-1}$~d$^{-1}$)\\
\hline

U, copper &  0.436 &    997.3& $<$1.0 10$^{-6}$ &$<$1.2 10$^{-3}$ & $<$8.9 10$^{-7}$ & $<$1.7 10$^{-6}$\\

Th, copper & 0.030  &997.3 &$<$1.0 10$^{-6}$ &$<$8.2 10$^{-5}$ & $<$6.2 10$^{-8}$& $<$1.2 10$^{-7}$ \\

U, PTFE & 26.2 & 3.96 &$<$5.0 10$^{-6}$ &$<$1.4 10$^{-3}$ &$<$1.1 10$^{-6}$& $<$2.1 10$^{-6}$\\

Th, PTFE &11.23  &3.96 &$<$10 10$^{-6}$ &$<$1.2 10$^{-3}$ &$<$9.1 10$^{-7}$ & $<$1.8 10$^{-6}$\\

U, steel &0.587 &920&2.6 10$^{-3}$ &3.8 &4.8 10$^{-4}$& 9.1 10$^{-4}$\\

Th, steel &0.189  &920 &1.1 10$^{-3}$ &5.5 10$^{-1}$ & 6.9 10$^{-5}$ & 1.3 10$^{-4}$\\

U, polyethylene &0.894 &3055 &$<$5.0 10$^{-6}$ &$<$3.7 10$^{-2}$ & $<$4.7 10$^{-6}$& $<$8.9 10$^{-6}$\\

Th, polyethylene &0.193 & 3055& $<$5.0 10$^{-6}$ & $<$8.1 10$^{-3}$ & $<$1.0 10$^{-6}$& $<$1.9 10$^{-6}$\\ \hline





total&&&&& $<$5.6 10$^{-4}$ & $<$1.1 10$^{-3}$\\ \hline


\end{tabular}
\caption{Estimates of the expected counting rate in TREX-DM using Ar
or Ne from radiogenic neutrons produced by uranium and thorium presence in the main materials of the set-up.}
\label{tabn}
\end{table}
\end{landscape}

\section{Sensitivity prospects}
\label{sen}

\begin{figure}
\centering
\includegraphics[width=0.5\textwidth]{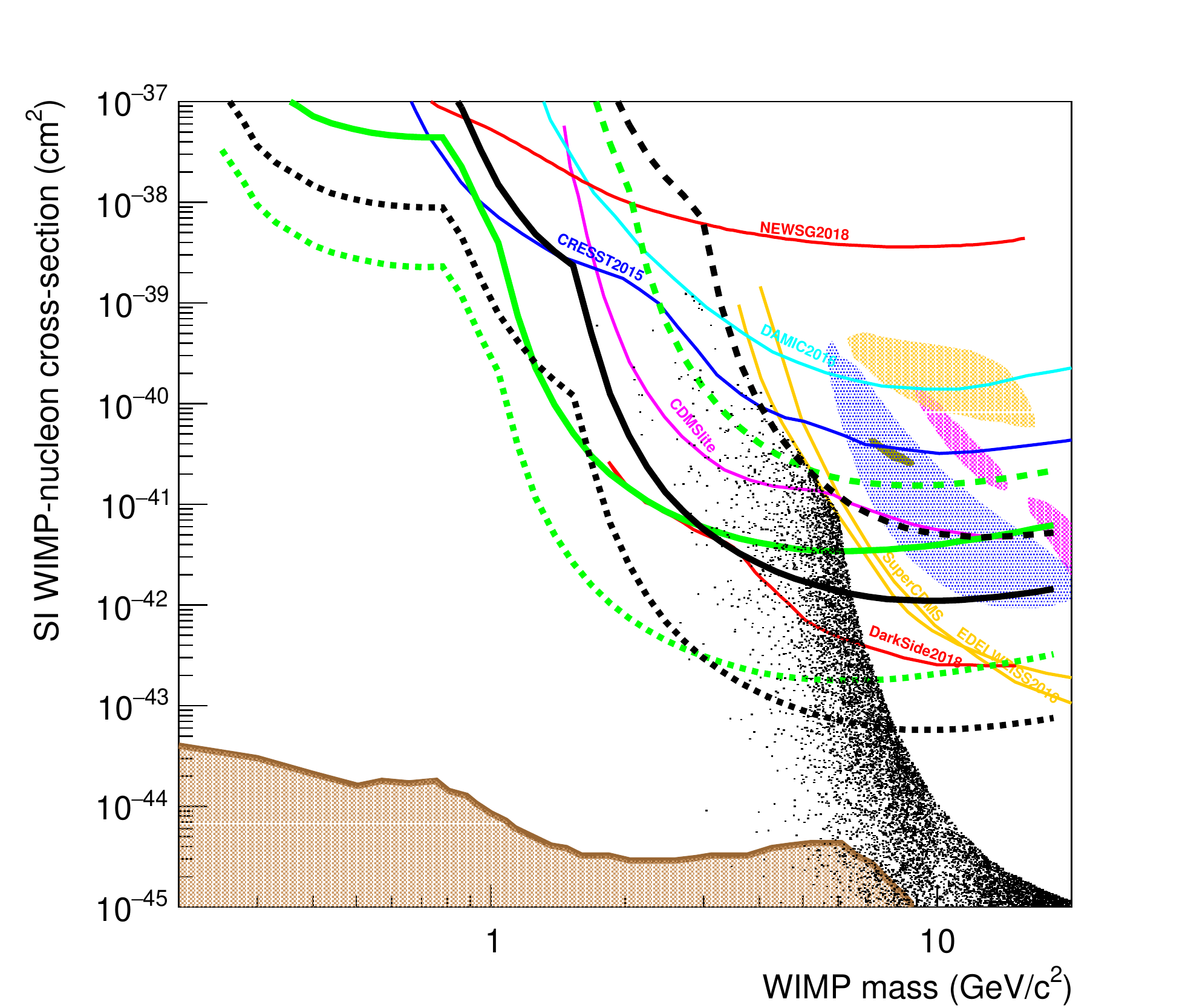} 
\caption{90\% C.L. sensitivity of TREX-DM under three different
conditions for background level, energy threshold and exposure,
summarized in table~\ref{scenarios}, for Ar+1\%iC$_{4}$H$_{10}$
(black lines) and Ne+2\%iC$_{4}$H$_{10}$ (green lines). The closed contours shown correspond to results from CDMS-II Si \cite{regCDMS} (blue, 90\% C.L.), CoGeNT \cite{rescogent} (dark gray, 90\% C.L.), CRESST-II \cite{regCRESST} (magenta, 95\% C.L.), and DAMA/LIBRA \cite{regDAMA} (yellow, 90\% C.L.). The brown shaded region corresponds to the sensitivity limit imposed by the solar neutrino coherent scattering background \cite{regneutrinos}. As a reference, 90\% C.L. exclusion limits from different experiments are also shown: CRESST-II 2016 \cite{rescresst} (dark blue), CDMSlite \cite{rescdms} (purple), DAMIC \cite{resdamic} (light blue), EDELWEISS \cite{resedelweiss} and SuperCDMS \cite{ressupercdms} (yellow) and DarkSide \cite{darkside18} and NEWS-G \cite{newsgresults} (red).}
\label{exclusion}
\end{figure}

For completeness, a first estimate of the achievable sensitivity in the search for low mass WIMPs is presented in this section in order to assess the effect of the background conditions evaluated and the expected detector performance. Figure~\ref{exclusion} presents the attainable exclusion plots (90\% C.L.) in the direct detection of WIMPs, for both Ar and Ne-based gas
mixtures at 10~bar, obtained assuming spin independent (SI)
interaction and standard values of the WIMP halo model with
Maxwell-Boltzmann velocity distribution and astrophysical parameters
(local dark matter density 0.3~GeV/c$^{2}$, local velocity 220~km/s,
laboratory velocity 232~km/s and galactic escape velocity 544~km/s).
The projected exclusion curves have been derived as in
\cite{iguaz16}, assuming the background is properly accounted for by
the background model and statistically subtracted. Three different
scenarios (labeled as A, B and C) for flat-shaped background, energy threshold and exposure
have been considered (see table~\ref{scenarios}). Despite the peak at $\sim$3~keV appearing in the RoI due to the $^{40}$K activity in the readout planes (see figure~\ref{figspectra}), the flat background assumption is reasonable as the exclusion plot is mainly determined by the lowest energy bins just above the energy threshold; in addition, works are underway to suppress this background contribution. A data-taking
campaign of approximately three years is foreseen, starting with Ne
with the option to change to depleted Ar. It can be seen that even in the first stage of TREX-DM with a small-scale set-up there is an option to explore regions of WIMPs uncovered by other experiments.

\begin{table}
\begin{center}
\begin{tabular}{llll}
\hline & A  &   B &  C   \\ \hline
Background level (counts keV$^{-1}$ kg$^{-1}$ d$^{-1}$) &  10&  1 &  0.1 \\
Energy threshold (keV$_{ee}$) &   0.4 & 0.1 & 0.1 \\
Exposure (kg y) & 0.3 & 0.3 & 10 \\ \hline
\end{tabular}
\end{center}
\caption{Conditions assumed in the calculations of the TREX-DM
sensitivity shown in figure~\ref{exclusion}.} \label{scenarios}
\end{table}

Following the rates summarized in table~\ref{rates}, the background assumed in scenario A gives the worst expectation. To achieve the level considered in scenario B, a reduction of the two main contributions from the quantified present activities should be enough. On one hand, as discussed in section \ref{secmicro}, work is underway to reduce the $^{40}$K content of the microbulk Micromegas readout and from the available results a reduction factor of $\sim$4 has been already achieved (see table \ref{res}). On the other hand, a new copper vessel produced using fresh copper and limiting exposure to cosmic rays at sea level to one month would have a $^{60}$Co activity of 0.01~mBq/kg (from the saturation activity deduced in~\cite{Heusser:2006}), giving a rate of 0.06(0.07)~counts keV$^{-1}$ kg$^{-1}$ d$^{-1}$ for Ar(Ne), which would mean a reduction by a factor 22 respect to the present copper vessel. The scenario B is interesting for Ne as for WIMP masses below 2~GeV/c$^{2}$ the sensitivity is beyond current bounds.

The background level considered in scenario C, predicting the stringent exclusion plots, appears as a plausible future goal. Several individual contributions are at the level of 0.1~counts keV$^{-1}$ kg$^{-1}$ d$^{-1}$: those of connectors, radon in air, muons and $^{39}$Ar for Ar. New connectors already screened (see results at \#19 of table~\ref{res}) would guarantee more than one order of magnitude of reduction for this contribution. Moreover, this background contribution could be totally suppressed if the new connection approach being tested is successful; multi-channel connectors can be replaced by a new custom-made ``face-to-face'' contact
technique offering in addition a substantial improvement in the level of connectivity. This system has been designed, manufactured, installed and it is presently being checked. The use of a muon veto system or a specific, more efficient, muon discrimination analysis together with the use of air from the radon-free air factory in operation in Canfranc offering an overall reduction factor of $\sim$1000 in the radon activity would also help to reduce these external contributions.

\section{Conclusions}
\label{con}

The TREX-DM experiment intends to look for
low mass WIMPs using a Micromegas-read High Pressure TPC filled with
Ar or Ne mixtures in the
Canfranc Underground Laboratory. At the beginning of 2019, it is at the commissioning phase and the data taking is expected to start soon using Ne+2\%iC$_{4}$H$_{10}$. Together with a sub-keV$_{ee}$ energy threshold,
an ultra-low background level at the lowest energy region is
mandatory; an assessment of the expected background has been
performed to help in the selection of radiopure components during
the design phase and to support a reliable estimate of the
experiment sensitivity. The background contributions, taking into
consideration all the known sources, have been simulated by means of
a dedicated Geant4 application and a custom-made code implementing
the detector response.

A material radioassay campaign has been carried out, based on
germanium spectrometry at LSC and complementary measurements, to
quantify the activity of all the relevant elements in the experiment
set-up (see table~\ref{res}). The total expected background level
from the internal activity should be below 6.1(6.6)~counts keV$^{-1}$
kg$^{-1}$ d$^{-1}$ for Ar(Ne), as shown in table~\ref{rates}; from that, 70(64)\% come from activities actually quantified. One of the
largest contributions is due to the copper vessel, cosmogenically
activated after being a few years at sea level, as shown in a
dedicated germanium measurement. This important contribution could
be suppressed by constructing a new vessel. The measured $^{40}$K
activity in the Micromegas readout gives also a significant rate and
for this reason new treatments are being analyzed in an
attempt to reduce this activity. It must be noted that the use of
underground argon has been assumed, considering the $^{39}$Ar
activity measured by DarkSide \cite{argon}. It has been verified
that a saturation activity of tritium in the gas media could be very
relevant, but the gas purification and obtention from underground
sources will avoid in principle this contribution. The effect of
radon and radon-induced activity on copper surfaces has been
assessed, finding it non-dominant at the present phase. The contribution from muons and
environmental neutrons is under control in the simulated conditions,
thanks to the background rejection capabilities and the shielding.
All in all, in the presently assumed conditions, the TREX-DM expected background can be considered
between 1 and 10~counts keV$^{-1}$ kg$^{-1}$ d$^{-1}$; a few improvements have been identified and are being tested in an attempt to further decrease it down to 0.1~counts keV$^{-1}$ kg$^{-1}$ d$^{-1}$, which would provide
a competitive sensitivity in the direct detection of low mass WIMPs.

\begin{acknowledgements}
This work has been financially supported by the European Commission
under the European Research Council T-REX Starting Grant ref.
ERC-2009-StG-240054 of the IDEAS program of the 7th EU Framework
Program and by the Spanish Ministry of Economy and Competitiveness
(MINECO) under Grants FPA2013-41085-P and FPA2016-76978-C3-1-P. We thank Rui de Oliveira from the EP-DT-DD Micro-Pattern Technologies (MPT) service at CERN for the preparation and cleaning of the Micromegas samples and the CEA/Saclay collaborators for their assistance. We also acknowledge LSC and GIFNA staff for their support.
\end{acknowledgements}



\end{document}